%% file: draft.tex
\documentclass[a4paper,10pt]{article}
\pdfoutput=1
\usepackage{jheppub}
\usepackage{slashed}
\usepackage[dvipsnames,table]{xcolor}
\usepackage{hyperref} 
\usepackage{xspace}
\usepackage[tight]{subfigure}
\usepackage{amsmath}
\usepackage{amssymb}
\usepackage{amsfonts}
\usepackage{mathrsfs}
\usepackage{comment}
\usepackage{afterpage}
\usepackage{verbatim}
\usepackage{booktabs}
\usepackage{array}
\usepackage[title,titletoc]{appendix}
\usepackage{tabularx}
\newcolumntype{C}[1]{>{\centering\arraybackslash}p{#1}}

\input{macros}

\title{From angular coefficients to quantum observables: ~~~~~~ a phenomenological appraisal in di-boson systems} 
\author{Michele Grossi,$^{a}$}
\affiliation[]{$^{a}$European Organisation for Nuclear Research (CERN), Espl. des Particules 1211 Geneva 23, Switzerland}
\author{Giovanni Pelliccioli,$^{b}$}
\affiliation[]{$^{b}$Max-Planck-Institut f\"ur Physik, Bolzmannstra{\ss}e 8, Garching, Germany}
\author{Alessandro Vicini\,$^{c}$}
\affiliation[]{$^{c}$Dipartimento di Fisica ``Aldo Pontremoli'', Universit\`a degli Studi di Milano and INFN, Sezione di Milano, Via Celoria 16, 20133 Milano, Italy
}
\emailAdd{michele.grossi@cern.ch}
\emailAdd{gpellicc@mpp.mpg.de}
\emailAdd{alessandro.vicini@mi.infn.it}

\date{\draftdate}
\preprint{COMETA-2024-24, MPP-2024-183, TIF-UNIMI-2024-15}

\abstract{
  Motivated by the growing interest in accessing the
  spin structure of multi-boson processes and 
  in measuring quantum entanglement at high energies,
  we study polarisation and spin-correlation coefficients in di-boson systems.
  We show that higher-order corrections of QCD and electroweak type, off-shell modelling, and realistic effects such as fiducial selections and neutrino reconstruction are unavoidable to properly determine such coefficients, and consequently to provide a sound interpretation of observables sensitive to quantum entanglement and Bell-inequality violation.
  Our findings are based on a detailed phenomenological analysis of 
  boson pairs at the LHC, either in inclusive electroweak production or coming from Higgs-boson decays.
  }


\begin{document}

\strut\hfill

\maketitle

\section{Introduction}
Measuring polarisation and spin correlations in multi-boson systems in high-energy particle collisions is a key to access the intricate structure of the electroweak (EW) sector of the Standard Model (SM), and more in general the deepest properties of quantum mechanics (QM).
The Large Hadron Collider (LHC) provides a unique opportunity to achieve these challenging purposes, thanks to its broad physics programme.

The simplest multi-boson system with a non-trivial 
helicity structure is formed by two EW bosons. 
This leads to a tree-level spin-density matrix that embeds 8 independent polarisation coefficients for each boson, and 64 spin-correlation coefficients. 
Besides the fact that the actual value of these coefficients depends on the production mechanism (EW production, vector-boson scattering, gluon fusion, Higgs-boson decay etc.), accessing them in the LHC environment is hampered by the application of fiducial selection cuts on decay products, the inclusion of higher-order corrections to the production and decay mechanisms, the presence of neutrinos amongst decay products, the contamination from non-resonant effects. It is crucial to notice that such obstacles represent actual physical effects, and as such their nature is both experimental and theoretical. 

The current paradigm for polarisation measurements of di-boson inclusive production \cite{ATLAS:2019bsc,CMS:2021icx,ATLAS:2022oge,ATLAS:2023zrv,ATLAS:2024qbd} and scattering \cite{CMS:2020etf} with the LHC Run-2 dataset is the so-called \emph{polarised-template method}, which is based on SM predictions for di-boson processes with 
definite polarisation states for intermediate EW bosons \cite{Ballestrero:2017bxn,BuarqueFranzosi:2019boy,Ballestrero:2019qoy,Ballestrero:2020qgv,Denner:2020bcz,Denner:2020eck,Poncelet:2021jmj,Denner:2021csi,Le:2022lrp,Le:2022ppa,Denner:2022riz,Denner:2023ehn,Dao:2023pkl,Hoppe:2023uux,Pelliccioli:2023zpd,Javurkova:2024bwa,Denner:2024tlu,Dao:2024ffg}.
This approach is somewhat complementary to the one employed by ATLAS and CMS collaborations for the Run-1 dataset, which relied on the extraction of \emph{angular coefficients} upon a kinematic extrapolation to the fully inclusive boson-decay phase space in single-boson production associated with jets \cite{CMS:2011kaj,ATLAS:2012au,CMS:2015cyj,ATLAS:2016rnf}, and in top-quark decays \cite{CMS:2016asd,CMS:2020ezf}.
The latter strategy, often dubbed \emph{quantum tomography}, is known to the community since long ago and is still developed by theory groups \cite{Boudjema:2009fz,Bern:2011ie,Stirling:2012zt,Aguilar-Saavedra:2015yza,Aguilar-Saavedra:2017zkn,Baglio:2018rcu,Baglio:2019nmc,Rahaman:2018ujg,Rahaman:2019lab,Frederix:2020nyw,Rahaman:2021fcz}. Interestingly, it has been recently revived in the lights of the growing interest in probing quantum entanglement and Bell inequalities at the LHC \cite{Barr:2021zcp,Aguilar-Saavedra:2022mpg,Aguilar-Saavedra:2022wam,Barr:2022wyq,Fabbrichesi:2023cev,Bernal:2023ruk,Fabbrichesi:2023jep,Fabbri:2023ncz,Aoude:2023hxv,Morales:2023gow,Bi:2023uop,Aguilar-Saavedra:2024whi,Bernal:2024xhm}. 

The study of entanglement probes and Bell-inequality tests in a di-boson system relies on the extraction from LHC data of the di-boson spin-density matrix, from a four-fermion final state. In fact, the quantum witnesses that are typically studied in the literature are combinations of the entries of the spin-density matrix, which eventually leave their footprints in the angular structure of the decay products of the two bosons.
All in all, measuring quantum entanglement at the LHC relies on the extraction of angular coefficients through quantum tomography, which in turn allows the reconstruction of the spin-density matrix. While this method has been first introduced relying on the tree-level structure of on-shell di-boson production and decay and assumes fully inclusive setups, a detailed assessment of the effects of fiducial cuts, higher-order corrections and off-shell modelling is still missing in the literature.

With this work, we investigate the aforementioned realistic effects, narrowing the gap with previous phenomenological studies and broadening the theoretical understanding needed for upcoming experimental studies in the field. 
The article is organised as follows. 
In Sect.~\ref{sec:compframe} we setup our computational framework for the extraction of angular coefficients, introduce spin correlations and their connection to quantum observables.
In Sect.~\ref{sec:procdef} we describe in detail the scattering and decay processes considered in this work, along with the corresponding numerical setups and the used simulation tools.
In Sect.~\ref{sec:results} we show the results of our numerical study and scrutinise the effect of off-shell modeling, higher-order corrections and selection cut on the extraction at the LHC of the angular coefficients that serve as inputs to relevant quantum observables. We draw our conclusions in Sect.~\ref{sec:conclus}.

\section{Computational framework}\label{sec:compframe}
\subsection{Angular coefficients}\label{sec:AngCoeff}

It is well known \cite{Bern:2011ie,Stirling:2012zt} that the decay rate of a weak boson into a pair of fermions, differential with respect to the decay angles, embeds information regarding the helicity state of the boson;   
for a spin-1 state, the latter can be represented in terms of 
spherical harmonics up to rank-2 \cite{Aguilar-Saavedra:2017zkn,Ballestrero:2017bxn,Baglio:2018rcu,Rahaman:2021fcz}.
Since it is relevant to identify the charge of the fermion to discriminate between left- and right-handed modes, we only focus on leptonic decays, namely,
\beq
\PZ\rightarrow \ell^+\ell^-, \qquad    \PW^+ \rightarrow \ell^+\nu_\ell\,,
\eeq
and identify the decay angles in the boson rest frame ($\theta,\phi$) as those associated to the positively charged lepton. The structure of a $\PW^-\rightarrow \ell^-\bar{\nu}_\ell$ can be easily obtained from the one of a $\PW^+$ by charge conjugation.
In formulas,
\begin{eqnarray}
\label{eq:V2ll}
\frac{\rd\sigma}{\rd\cos\theta \,\rd\phi \,\rd \mc X} 
&=& 
\frac{\rd \sigma}{\rd \mc X}
\left[\frac1{4\pi}+
\sum_{l=1}^2 \sum_{m=-l}^l\,\alpha_{lm}(\mc X)\,Y_{lm}(\theta,\phi)
\right]\,,
\end{eqnarray}
where $\mc X$ is any other kinematic observable independent of the decay angles, \eg the transverse momentum of the decaying boson.
The $Y_{lm}(\theta,\phi)$ are real spherical harmonics, normalised in such a way that 
\beq
\int_{-1}^{1} \!\rd\cos\theta\int_{0}^{2\pi}\!\rd\phi \,\,Y_{lm}(\theta,\phi)\,Y_{l'm'}(\theta,\phi) = \delta_{ll'}\delta_{mm'}\,.
\eeq 
When no selection cuts are applied on individual decay products, the 8 angular coefficients can be easily extracted according to
\begin{equation}\label{eq:inc}
    \frac{\rd \sigma}{\rd \mc X}\,\alpha_{lm}(\mc X)\,\equiv\,
    \int_{-1}^{1} \!\rd\cos\theta\int_{0}^{2\pi}\!\rd\phi\,\,\frac{\rd\sigma}{\rd\phi 
    \,\rd\cos\theta\,\rd \mc X} Y_{lm}(\theta,\phi) \, .
\end{equation}
We notice that the expansion in spherical harmonics is equivalent to the one in other projectors \cite{Bern:2011ie,Stirling:2012zt,Ballestrero:2017bxn,Baglio:2018rcu,Baglio:2019nmc} or asymmetries \cite{Rahaman:2018ujg,Rahaman:2019lab,Rahaman:2021fcz}, provided that they represent a complete basis for the $\ell\leq2$ angular structure.
The extension of Eq.~\ref{eq:V2ll} to the case of two weak bosons decaying into pairs of leptons,
\beq
V (\rightarrow \ell_1 \ell_2) V' (\rightarrow \ell_3 \ell_4)
\eeq
can be written
\cite{Rahaman:2021fcz,Aguilar-Saavedra:2022wam} as follows,
\begin{eqnarray}
\label{eq:V1V2llll}
\frac{\rd\sigma}{\rd\cos\theta_1 \,\rd\phi_1\,\rd\cos\theta_3 \,\rd\phi_3 \,\rd \mc X} 
&=& 
\frac{\rd \sigma}{\rd \mc X}
\bigg[
\frac1{(4\pi)^2}+\\
&&\hspace*{-2cm}+\frac1{4\pi}\,\sum_{l=1}^2 \sum_{m=-l}^{l}\,\alpha^{(1)}_{lm}(\mc X)\,Y_{lm}(\theta_1,\phi_1)
\nnb\\
&&\hspace*{-2cm}
+\frac1{4\pi}\,\sum_{l=1}^2 \sum_{m=-l}^{l}\,\alpha^{(3)}_{lm}(\mc X)\,Y_{lm}(\theta_3,\phi_3)
\nnb\\
&&\hspace*{-2cm}
+\sum_{l_1=1}^2\sum_{l_3=1}^2 \sum_{m_1=-l_1}^{l_1}
  \sum_{m_3=-l_3}^{l_3}\,\gamma_{l_1m_1l_3m_3}(\mc X)\,Y_{l_1m_1}(\theta_1,\phi_1)\,Y_{l_3m_3}(\theta_3,\phi_3)
\bigg]\,,\nnb
\end{eqnarray}
where $\theta_1,\phi_1$ ($\theta_3,\phi_3$) are the polar and azimuthal angles associated to the decay of the first (second) boson. Notice that $\theta_1,\phi_1$ are computed in the rest frame of the $(\ell_1,\ell_2)$ system, while $\theta_3,\phi_3$ are computed in the rest frame of the $(\ell_3,\ell_4)$ system.

\begin{figure}[t]
  \centering
    \includegraphics[scale=0.33]{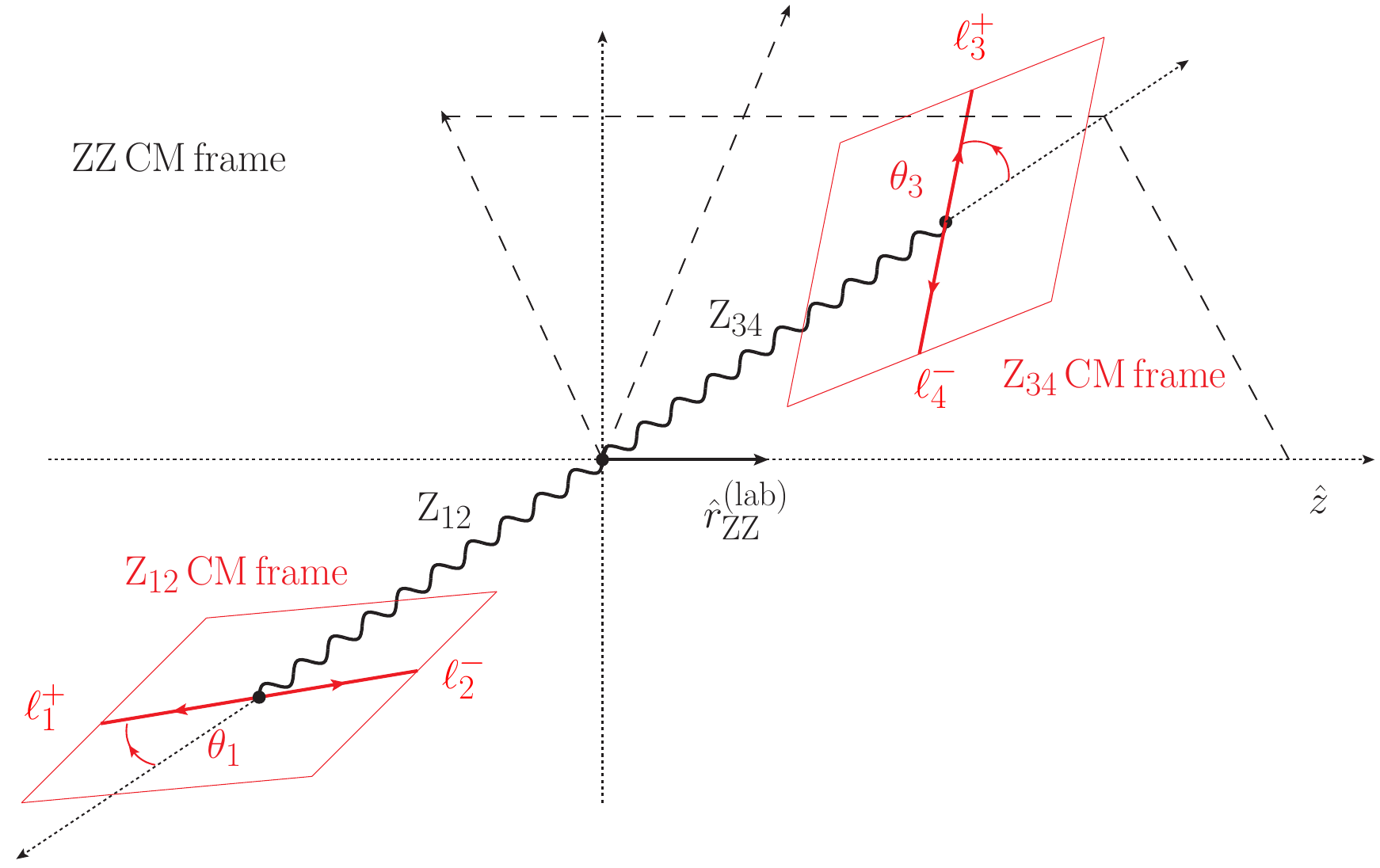}   
  \caption{Definition of the coordinate system in the case of a $\PZ\PZ$ system.    }\label{fig:geomscheme}
\end{figure}

As a last remark of this subsection, we recall that while the analytic structure of Eq.~\ref{eq:V1V2llll} (and of Eq.~\ref{eq:V2ll}) remains the same in any helicity reference frame, the actual value of the extracted coefficients depends on the choice of the coordinate system and especially on the Lorentz frame where the helicity states of intermediate bosons are defined, \emph{i.e.} the reference frame for the spin quantisation.
The di-boson system is identified as the system formed by the four \emph{physical} leptons; such a choice is robust in the presence of higher-order corrections.
For di-boson systems the typical frames chosen for polarisation and spin-correlation studies are the laboratory, and the di-boson centre-of-mass (CM) frame. 
The latter, at LO, coincides with the partonic CM frame in EW production or with the Higgs-boson rest frame in Higgs decays.
The studies of quantum observables for entanglement and Bell-inequality tests
\cite{Barr:2021zcp,Aguilar-Saavedra:2022mpg,Aguilar-Saavedra:2022wam,Barr:2022wyq,Fabbrichesi:2023cev,Bernal:2023ruk,Fabbrichesi:2023jep,Fabbri:2023ncz,Aoude:2023hxv,Morales:2023gow,Bi:2023uop,Aguilar-Saavedra:2024whi,Bernal:2024xhm} typically employ the di-boson CM frame.
In this work, unless otherwise stated, we will employ the di-boson CM frame as Lorentz frame for the helicity-state definition, and the di-boson-system direction in the laboratory as a reference axis to determine azimuthal decay angles.
An illustration of the chosen reference frame, often dubbed \emph{modified helicity coordinate system} \cite{ATLAS:2019bsc}, is given in Fig.~\ref{fig:geomscheme}.
In our investigation of Higgs-boson decays in Sect.~\ref{sec:HiggsHO}, we will compare this choice with the one of the Higgs-boson rest frame.

\subsection{Polar spin correlations and the $R_c$ quantity}\label{subsec:Rc}
The full structure of Eq.~\ref{eq:V1V2llll} 
contains relevant information about how correlated the spin of the two EW bosons are. In other words, the difference between Eq.~\ref{eq:V1V2llll} and a factorised form of it resulting from the combination of two Eqs.~\ref{eq:V2ll} (one for each boson) gives a number of quantitative measures of the spin correlations. 
In order to give an example, we consider the simple case of a quantity that can be constructed uniquely with polar angular coefficients.
Integrating over the azimuthal angles, the decay rate differential in the two polar decay angles reads,
\beqn\label{eq:coscos}
\frac{\rd\sigma}{\rd\cos\theta_1 \,\rd\cos\theta_3 \,\rd \mc X} 
&=& 
\frac{\rd \sigma}{\rd \mc X}
\bigg[
\frac14+\pi\,\sum_{l=1}^2 \,\alpha^{(1)}_{l,0}(\mc X)\,Y_{l,0}(\theta_1)+\pi\,\sum_{l=1}^2 \alpha^{(3)}_{l,0}(\mc X)\,Y_{l,0}(\theta_3)\\
&&\hspace*{0.6cm}+\,4\pi^2\,\sum_{l_1=1}^2\sum_{l_3=1}^2 
  \,\gamma_{l_1,0,l_3,0}(\mc X)\,Y_{l_1,0}(\theta_1)\,Y_{l_3,0}(\theta_3)
\bigg]\,,\nnb
\eeqn
where $Y_{l_i,0}(\theta_i) = Y_{l_i,0}(\theta_i,\phi_i=0)$ is a Legendre polynomial of degree $l_i$, up to a normalisation factor. 
Eq.~\ref{eq:coscos} embeds 8 independent coefficients and can be re-written in terms of the joint polarisation fractions $f_{\lambda\lambda'}\,$(first polarization index for the first boson, second index for the second boson), with $\lambda',\lambda=\rL$ (longitudinal), $+$ (right handed), $-$ (left handed):
\beqn\label{eq:jointpolfrac}
\frac{\rd\sigma}{\rd\cos\theta_1 \,\rd\cos\theta_3 \,\rd \mc X} =
\frac{9}{64}\,\frac{\rd\sigma}{\rd \mc X} &\Big[&4\,f_{\rL\rL}(\mc X) \sin ^2\theta_1\,\sin ^2\theta _3 \\
&+&2\, f_{\rL-}(\mc X)\sin ^2\theta_1 \left(1-2 \eta _3 \cos \theta_3+\cos ^2\theta_3\right)\nnb\\[0.12cm]
&+&2\, f_{\rL+}(\mc X) \sin ^2\theta_1 \left(1+2 \eta _3 \cos \theta_3+\cos   ^2\theta_3\right)\nnb\\[0.12cm]
&+&2\, f_{-\rL}(\mc X) \left(1-2 \eta _1 \cos \theta_1+\cos ^2\theta_1\right)\sin ^2\theta_3 \nnb\\[0.12cm]
&+&2\, f_{+\rL}(\mc X) \left(1+2 \eta _1 \cos \theta_1+\cos^2\theta_1\right) \sin ^2\theta_3 \nnb\\[0.12cm]
&+& f_{--}(\mc X) \left(1-2 \eta _1 \cos \theta_1+\cos ^2\theta_1\right) \left(1-2 \eta _3 \cos\theta_3+\cos ^2\theta_3\right)\nnb\\[0.12cm]
&+& f_{-+}(\mc X) \left(1-2 \eta _1 \cos \theta_1+\cos ^2\theta_1\right) \left(1+2   \eta _3 \cos \theta_3+\cos ^2\theta_3\right)\nnb\\[0.12cm]
&+& f_{+-}(\mc X) \left(1+2 \eta _1 \cos \theta_1+\cos ^2\theta_1\right) \left(1-2   \eta _3 \cos \theta_3+\cos ^2\theta_3\right)\nnb\\
&+& f_{++}(\mc X) \left(1+2 \eta _1 \cos \theta_1+\cos ^2\theta_1\right) \left(1+2 \eta _3 \cos \theta_3+\cos ^2\theta_3\right)\,\Big]\,,\nnb
\eeqn
where the parameter $\eta_\ell$ is the asymmetry between the left- and right-chirality coupling strengths of the EW-boson to leptons ($\eta_\ell=\eta_1=\eta_3$). This quantity can be expressed, in the $(G_\mu,\MW,\MZ)$ input scheme, for massless leptons, as
\beq
\eta_\ell = \frac{c_{\ell,-}^2-c_{\ell,+}^2}{c_{\ell,+}^2+c_{\ell,-}^2} = 
\frac{4 \text{\MW}^2 \text{\MZ}^2-3 \text{\MZ}^4}{8 \text{\MW}^4-12 \text{\MW}^2 \text{\MZ}^2+5 \text{\MZ}^4} 
\approx
0.219\,.
\eeq
The joint polarisation fractions in Eq.~\ref{eq:jointpolfrac} sum up to unity and are related to single-boson polarisation fractions $f^{(1)}_{\lambda}$ (first boson) and $f^{(3)}_{\lambda'}$ (second boson) by means of the relations,
\beqn\label{eq:singlejoint}
\sum_{\lambda,\lambda'}f_{\lambda \lambda'} = 1\,,\quad
\sum_{\lambda'}f_{\lambda\lambda'}=f^{(1)}_{\lambda}\,,\quad
\sum_{\lambda}f_{\lambda\lambda'}=f^{(3)}_{\lambda'}\,.
\eeqn
Using Eq.~\ref{eq:singlejoint}, the independent coefficients in Eq.~\ref{eq:coscos} are easily expressed as linear combinations of single-boson and joint polarisation fractions: 
\beqn\label{eq:alphastofraction}
\alpha^{(i)}_{1,0}    &=& \frac14\sqrt{\frac3{\pi}}\,\eta_i\left(f^{(i)}_{+}-f^{(i)}_{-}\right)\,,\quad
\hspace{2.5cm}\alpha^{(i)}_{2,0} \,=\,\frac{1-3\,f^{(i)}_{\rL}}{4\sqrt{5\pi}}\,,\\
\gamma_{1,0,1,0} &=&
3\,\eta_1 \eta_3\, \frac{f_{--}+f_{++}-f_{-+}-f_{+-}}{16\pi}\,,\quad 
\hspace{0.9cm}
\gamma_{2,0,2,0} \,=\,
\frac{1-3f^{(1)}_{\rL}-3f^{(3)}_{\rL}+9f_{\rL\rL}}{80\pi}\,,
\nnb\\
\gamma_{1,0,2,0} &=&
\sqrt{\frac35}\,\eta_1\,\frac{
3
\left( 
f^{\,}_{-\rL}-f^{\,}_{+\rL}
\right)
-
\left( 
f^{(1)}_{-}-f^{(1)}_{+}
\right)
}{16\pi}\,,\quad
\gamma_{2,0,1,0} \,=\,
\sqrt{\frac35}\,\eta_3\,\frac{
3
\left( 
f^{\,}_{\rL-}-f^{\,}_{\rL+}
\right)
-
\left( 
f^{(3)}_{-}-f^{(3)}_{+}
\right)
}{16\pi}\,.\nnb
\eeqn
It has been recently proposed by ATLAS \cite{ATLAS:2022oge} to evaluate the level of correlation between the two longitudinally polarised bosons through the
variable $R_c$, which is directly related to angular coefficients with $l=2,\,m=0$,
\beq\label{eq:Rcdef}
R_c=\frac{f_{\rL\rL}}{f^{(1)}_{\rL} \,f^{(3)}_{\rL}}
=\frac{1-4\sqrt{5\pi} \,\alpha^{(1)}_{2,0}-4\sqrt{5\pi} \,\alpha^{(3)}_{2,0}+80\pi\,\gamma_{2,0,2,0}}{1-4\sqrt{5\pi} \,\alpha^{(1)}_{2,0}-4\sqrt{5\pi} \,\alpha^{(3)}_{2,0}+80\pi\,\alpha^{(1)}_{2,0}\,\alpha^{(3)}_{2,0}}\,,
\eeq
where the second expression is easily obtained inverting the relations in Eq.~\ref{eq:alphastofraction}.
If the spin states of the two bosons were completely uncorrelated, then one would have $R_c=1$. A value different from 1 has been measured by ATLAS in $\PZ\PW$ inclusive production \cite{ATLAS:2022oge} and is predicted in the SM at NLO QCD+EW \cite{Denner:2020eck,Le:2022lrp}, clearly pointing in a marked correlation.
A word of caution is needed here. While the quantity $R_c$ can be evaluated in any kinematic setup as a ratio of polarisation fractions, thanks to the polarisation-template method \cite{Ballestrero:2017bxn,BuarqueFranzosi:2019boy,Ballestrero:2019qoy,Ballestrero:2020qgv,Denner:2020bcz,Denner:2020eck,Poncelet:2021jmj,Denner:2021csi,Le:2022lrp,Le:2022ppa,Denner:2022riz,Denner:2023ehn,Dao:2023pkl,Hoppe:2023uux,Pelliccioli:2023zpd,Javurkova:2024bwa,Denner:2024tlu,Dao:2024ffg},
its evaluation in terms of the angular coefficients extracted with projections (see Eq.~\ref{eq:inc}) only makes sense in a fully inclusive setup, namely in the absence of transverse-momentum and rapidity cuts on the decay products. A detailed discussion of this aspect is carried out in Sect.~\ref{sec:selectioncuts}.

\subsection{Quantum observables}\label{sec:quantumobs}
The characterisation of the spin structure of di-boson systems is being increasingly explored nowadays in the context of quantum-information observables, under the compound definition of \emph{quantum observables}. Concurrence bounds, purity, and Bell inequalities are studied for bipartite systems of qubits or qutrits, the latter representing two fermions or massive gauge bosons respectively. 
Quantum observables at TeV-scale colliders like the LHC do not only serve as further probes of the SM structure, but also as complementary probes for potential new-physics effects, parameterised for example in the Standard Model Effective Field Theory (SMEFT). 

The development of a direct connection between high-energy physics and quantum information from a theoretical viewpoint is mature enough to consider its effective integration in experimental analyses at colliders. 
However, the interpretation of the spin-density matrix 
in terms of general quantum properties of LHC processes
requires some considerations about the role of higher-order
effects and selection cuts, which are unavoidable in a 
collider environment.
This is the purpose of this work, but before going into the details 
it is worth recalling some general notation which strictly pertains
quantum information and making the connection to physical objects 
that can be actually accessed in LHC events. 

Following the general formalism of Ref.~\cite{Aguilar-Saavedra:2015yza,Aguilar-Saavedra:2017zkn,Aguilar-Saavedra:2022wam}, 
the explicit structure of Eq.~\ref{eq:V1V2llll}, already written in terms of angular decay variables,
comes from the formal expression,
\beq\label{eq:formal}
    \frac{1}{\sigma}\frac{d\sigma}{d\Omega_1d\Omega_2} = \left (\frac{3}{4\pi}\right)^2 {\rm Tr}~\{\rho\,(\Gamma^{(1)} \otimes \Gamma^{(3)})^{\rT}\}\,.
\eeq
In Eq.~\ref{eq:formal}, $\rho$ is the hermitian $9\times 9$ spin-density matrix for a boson-pair system,
\begin{equation}
\label{eq:spindm}
    \rho = \frac{1}{9} \left[ \left({\mathbb{I}}_3 \otimes  {\mathbb{I}}_3\right) + \sum_{l,m}\,A^{(1)}_{l,m} \left(T^l_m \otimes {\mathbb{I}}_3\right)  + \sum_{l,m}\,A^{(3)}_{l,m} \left({\mathbb{I}}_3 \otimes T^l_m\right) + C_{l_1,m_1,l_3,m_3} \left(T^{l_1}_{m_1} \otimes T^{l_3}_{m_3}\right)\right]\,,
\end{equation}
where $\mathbb{I}_3$ is the identity and $T^l_m$ are irreducible tensor representations of the vector-boson spin \cite{Aguilar-Saavedra:2022wam}, while $A^{(1,3)}_{l,m}$ and $C_{l_1,m_1,l_3,m_3}$ are complex-valued coefficients. The $\rho$ matrix is multiplied by the tensor product $\Gamma^{(1)} \otimes \Gamma^{(3)}$, where $\Gamma$ is the density matrix associated to a spin-1-boson decay into two spin-half fermions.
The correspondence between Eq.~\ref{eq:formal} and Eq.~\ref{eq:V1V2llll} is then evident, since
\beq
{\rm Tr}\left( {\mathbb I}_3 \Gamma^{\rT}\right) \propto Y_{00}(\theta,\phi)\,,\qquad 
{\rm Tr}\left( T^l_m \Gamma^{\rT}\right) \propto Y_{lm}(\theta,\phi)\,,
\eeq
up to suitable real multiplicative factors.
More precisely, the $\alpha,\gamma$ coefficients defined in Eq.~\ref{eq:V1V2llll} can be converted into the $A,C$ coefficients of \citere{Aguilar-Saavedra:2022wam} by means of the following relations:
\beq
\sqrt{40\pi}\alpha^{(i)}_{2,m_i} = A^{(i)}_{2,m_i}\,,\qquad 
40\pi\gamma_{2,m_1,2,m_3} = C_{2,m_1,2,m_3}\,,\qquad
8\pi\gamma_{1,m_1,1,m_3}/\eta_\ell^2 = C_{1,m_1,1,m_3}\,.
\eeq
Notice also that the decay angles defined in this work are associated to positively charged leptons, while in \citere{Aguilar-Saavedra:2022wam} are associated to the negatively charged leptons.

Performing a quantum-state tomography of the global system, \emph{i.e.} the four-lepton system in di-boson, allows to obtain information about its quantum properties.
A general procedure to probe the separability of a spin-density matrix, \emph{i.e.} if the two bosons are entangled or not, is not known. However, in the case of massive-boson pairs (two qutrits) the Peres-Horodecki condition \cite{Simon:1999lfr} is sufficient for entanglement. If the di-boson system comes from the decay of a scalar particle, sufficient and necessary conditions for the two bosons to be entangled are:
\begin{equation}\label{eq:entangcond}
    \gamma_{2,1,2,-1} \neq 0\,\qquad \textrm{or} \qquad \gamma_{2,2,2,-2} \neq 0\,.
\end{equation}
Furthermore, if a quantum state is close to maximal entanglement one can also probe Bell non-locality.
A suitable Bell inequality for two-qutrits systems is the CGLMP one \cite{Collins:2002sun}, which
can be expressed in terms of the expectation value of a Bell operator, 
\begin{equation}
\label{eq:CGLMP}
    I_3 = \langle \mathcal{O}_{\rm Bell}\rangle = {\rm Tr}\left(\rho \,\mathcal{O}_{\rm Bell} \right) \leq 2\,.
\end{equation}
For the case of a scalar decaying into two spin-1 massive bosons, a Bell operator that leads to a maximal violation of Eq.~\ref{eq:CGLMP} is given by \cite{Aguilar-Saavedra:2022wam},
\begin{eqnarray}
    {\mc O}_{\rm  Bell}&=& \frac{2}{3\sqrt{3}} (T_1^1 \otimes T_1^1 - T_0^1 \otimes T_0^1  + T_1^1 \otimes T_{-1}^1 ) +\frac{1}{12} ( T_2^2\otimes T_2^2 + T_2^2 \otimes T_{-2}^2 ) \\
    &+& \frac{1}{2\sqrt{6}} ( T_2^2\otimes T_0^2 + T_0^2 \otimes T_2^2 ) - \frac{1}{3} ( T_1^2\otimes T_1^2 + T_1^2 \otimes T_{-1}^2 ) + \frac{1}{4} T_0^2 \otimes T_0^2 \,,\nnb
     + {\rm h.c.}
\end{eqnarray}
where the same irreducible tensor representations introduced in Eq.~\ref{eq:spindm} are used.
Written according to the angular expansion of Eq.~\ref{eq:V1V2llll}, the explicit form of $I_3$ reads,
\beq\label{eq:I3explicit}
I_3\,=\, \frac12 + \frac{4\sqrt{3}}{9} - \sqrt{5\pi}\left( 1 - \frac{8\sqrt{3}}{9}\right)\,\alpha_{2,0}    - 40\pi\left(\frac23+\frac{4\sqrt {3}}{9}\right)\,\gamma_{2,1,2,-1} + \frac{20\pi}{3}\,\gamma_{2,2,2,-2} \,.
\eeq
Sufficient conditions for entanglement and Bell operators exist also for general boson-pair systems \cite{Fabbrichesi:2023cev}, but are not considered in this work.

While the connection between decay-angle coefficients and quantum observables is now made explicit, it is still based on a tree-level description of the considered processes and on the
possibility to extrapolate to a fully inclusive phase space.
In order to have a solid theoretical understanding, we raise the following issue: what is the validity limit of the tree-level angular expansion up to rank-2 spherical harmonics, given the 
presence of higher-rank contributions introduced by radiative effects and fiducial-cut application? Translated in terms of quantum observables: are tree-level expressions for 
such quantities (like Eq.~\ref{eq:I3explicit} for $I_3$) still suitable observables to
claim quantum entanglement or Bell-inequality violation, knowing that the defintion of
the spin-density matrix $\rho$ in Eq.~\ref{eq:spindm} is either incomplete or not even 
well defined?

In the following we address the above considerations by means of a detailed discussion of realistic effects which come into play at colliders.

\section{Process definition and numerical setup}\label{sec:procdef}
\subsection{Di-boson inclusive production at the LHC}
We consider the production at the LHC of two pairs of leptons belonging to different families, in order to avoid additional systematic uncertainties from identical leptons,
\beqn
\Pp\Pp\rightarrow &\Pe^+\Pe^-&\mu^+\mu^-\,+X\,,\qquad (\PZ\PZ)\nnb\\
\Pp\Pp\rightarrow &\Pe^+\nu_{\Pe}&\mu^+\mu^-\,+X\,.\qquad (\PW^+\PZ) \label{eq:VVdef}
\eeqn
We model the above processes in the pole-approximation approach described in \cite{Denner:2020bcz,Denner:2020eck,Denner:2021csi,Denner:2022riz,Denner:2023ehn,Pelliccioli:2023zpd} as well as in the full off-shell picture. 
Higher-order corrections are included up to NLO QCD and NLO EW accuracy, and complementing NLO QCD results with QCD+QED Parton-Shower (PS) effects.
The input SM parameters are the same as those used in  \citeres{Denner:2021csi,Denner:2020eck} and the events are generated with with \PB-{\sc Res} \cite{Chiesa:2020ttl,Pelliccioli:2023zpd} .
Two kinematic selections are considered for both processes. A first one, dubbed \emph{inclusive}, 
uniquely involves an invariant-mass cut on
charged-lepton pairs with opposite charge and same flavour,
\beq\label{eq:inclsetup}
81\GeV < M_{\ell^+\ell^-} < 101\GeV\,,\qquad \ell=\Pe,\mu\,.
\eeq
As second kinematic selections, we choose ATLAS \emph{fiducial} regions employed in recent polarisation measurements. On top of the invariant-mass cut in Eq.~\ref{eq:inclsetup}, for the ZZ process we apply \cite{ATLAS:2023zrv}:
\beq\label{eq:ATLASfid}
\pt{\Pe(\mu)} > 7 (5)\GeV, \quad|y_{\Pe(\mu)}|<2.47 (2.7),\quad \pt{\ell_{1(2)}}>20\GeV,\quad M_{4\ell}>180\GeV,\quad\Delta R_{\ell\ell'}>0.05\,,
\eeq
while for WZ we apply \cite{ATLAS:2022oge}:
\beq\label{eq:ATLASfidWZ}
\pt{\Pe(\mu)} > 15(20)\GeV, \quad|y_{\Pe(\mu)}|<2.5,\quad M_{\rm T, \PW}>30\GeV,\quad\Delta R_{\Pe^+\Pe^-}>0.2,\quad\Delta R_{\Pe^\pm\mu^+}>0.3\,.
\eeq
The dressing of charged leptons with additional photon radiation is carried out with resolution radius $R_{\ell\gamma}=0.1$. 
In the case of $\PW\PZ$ production, the presence of a single neutrino allows to reconstruct the four-lepton kinematics with standard on-shell-ness requirements \cite{ATLAS:2019bsc}.
While interesting from a theoretical point of view, the $\PW^+\PW^-$ mechanism is characterised by the presence of two neutrinos, whose separate reconstruction is not possible with standard techniques. Therefore, we do not show any result about it.

\subsection{Higgs-boson decay into four leptons}
In addition to di-boson inclusive production, we consider the decay of a SM Higgs-boson into four massless charged leptons, which is considered as one of the most promising processes to detect quantum entanglement \cite{Barr:2022wyq,Aguilar-Saavedra:2022wam}.
We consider the following decay reaction,
\beq\label{eq:Higgsdec}
{\rm H}\rightarrow \Pe^+\Pe^-\mu^+\mu^-\,(+\gamma)\,\qquad (\PZ\PZ^*)\,,
\eeq
including complete EW corrections at NLO. A Higgs-boson mass of $125\GeV$ is understood and a minimum-invariant-mass cut is applied on both opposite-sign, same-flavour lepton pairs,
\beq\label{eq:mll10}
M_{\ell^+\ell^-} > 10\GeV\,,\qquad \ell=\Pe,\mu\,.
\eeq
The lepton dressing is carried out with resolution radius $R_{\ell\gamma}=0.1$.
The numerical results at LO and NLO EW have been obtained with the
{\sc MG5\_aMC@NLO} event generator \cite{Alwall:2014hca,Frederix:2018nkq} and checked independently with {\sc Prophecy4f} \cite{Denner:2019fcr}.

\section{Results}\label{sec:results}
In this section we unravel the intricate effects of a realistic modeling of di-boson production and decay at hadron colliders in view of the extraction of polarisation and spin-correlation coefficients. 
In particular, we focus on how off-shell effects, higher-order corrections of both QCD and EW type, and selection cuts modify the extraction of entries of the spin-density matrix for one or more resonant particles. To give quantitative evidence to such effects we consider the processes defined in Eqs.~\ref{eq:VVdef} and \ref{eq:Higgsdec}.

\subsection{Off-shell vs on-shell modelling}\label{sec:offshell}
The final state relevant for the study  of di-boson production is given by two pairs of fermions, with the purely leptonic case preferable for the high experimental precision in the reconstruction of the boson kinematics.
The dominant contribution to the amplitude is given by those configurations with both bosons produced in the $s$-channel with a virtuality close to the EW-boson pole mass, which subsequently decay, and are dubbed as double resonant. A convenient proxy to analyze this subset of contributions is given by the amplitude for the production of two on-shell bosons, where the identification of the boson polarisation states is precise.

The comparison of kinematical distributions computed alternatively with the complete amplitude for the four-lepton final state, or 
 in the pole \cite{Denner:2020bcz,Denner:2020eck,Poncelet:2021jmj,Denner:2021csi,Denner:2023ehn} or in the narrow-width approximations \cite{Poncelet:2021jmj,Hoppe:2023uux}, 
 shows that off-shell effects have an impact by less than 2\% in most of phase-space regions, as expected from the intrinsic accuracy of the employed on-shell approximations \cite{Denner:2000bj,Uhlemann:2008pm}.

In the double resonant limit each of the two bosons can be described relying on the angular decomposition of Eqs.~\ref{eq:V2ll} and \ref{eq:V1V2llll}, which in the case of leptonic decays remains valid in the presence of higher-order QCD corrections.
The off-shell effects break this simple picture, with an impact on the value of the 80 coefficients needed to describe the angular distribution of the four leptons.

\begin{figure}[b]
  \centering
   \subfigure[LO\label{fig:LOdiag}]{\includegraphics[scale=0.35]{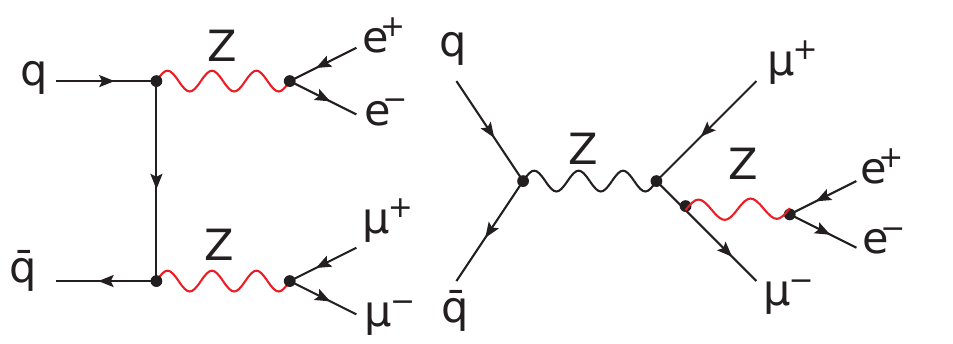}}
   \subfigure[NLO QCD\label{fig:NLOQCDdiag}]{\includegraphics[scale=0.35]{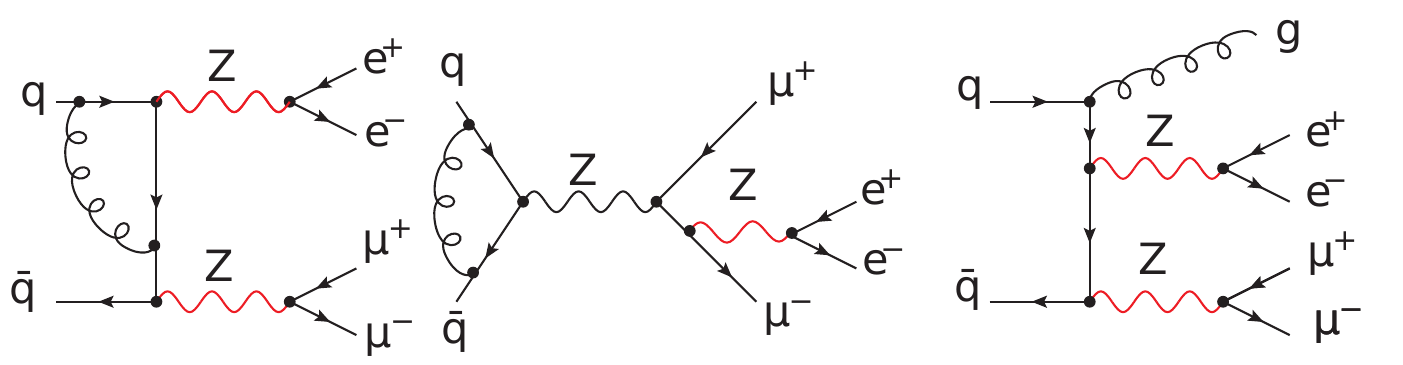}}\\
   \subfigure[NLO EW (virtual)\label{fig:NLOEWdiag}]{\includegraphics[scale=0.35]{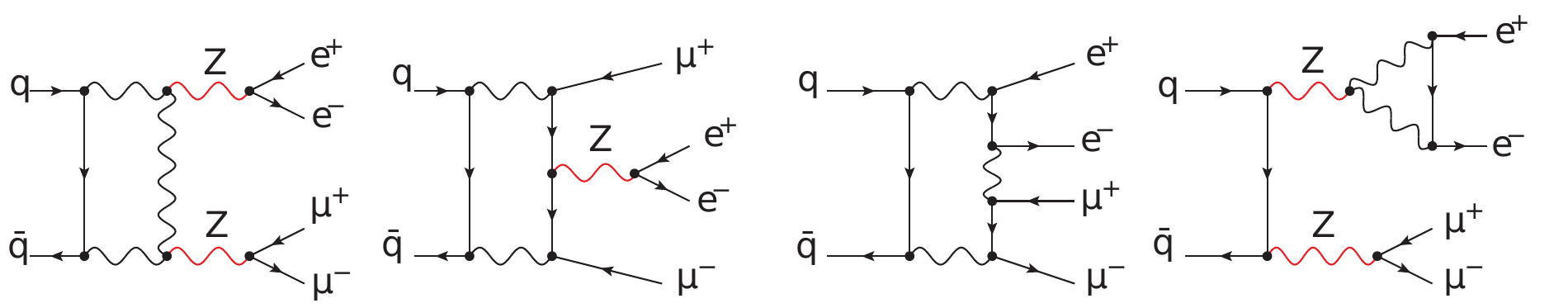}}\\
   \subfigure[NLO EW (real)\label{fig:NLOQEDLOdiag}]{\includegraphics[scale=0.35]{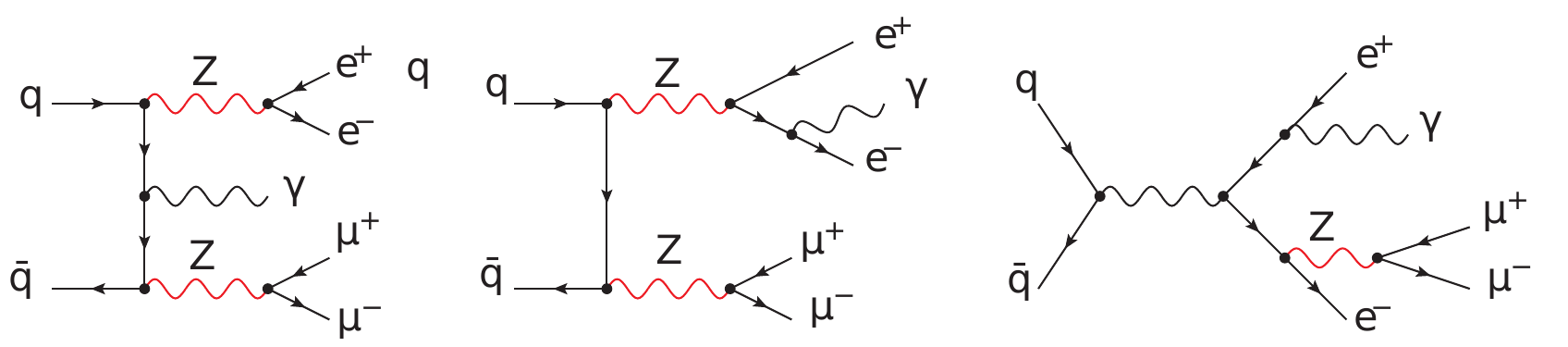}}
  \caption{Sample diagrams contributing to off-shell $\PZ\PZ$  production in the $q\bar{q}$ channel at tree level (a), at NLO QCD (b) and at NLO EW (c, d). Z bosons in $s$-channel and decaying into two same-flavour, opposite-sign leptons (and an additional photon at NLO EW) are marked in red. Resonant contributions entering the DPA calculation are those with two $\PZ$-boson propagators marked in red.  \label{fig:ZZdiagrams}}
\end{figure}

The first relevant aspect for the extraction of angular coefficient is the off-shell modeling of EW bosons. It is clear that the tree-level analytic structure of Eqs.~\ref{eq:V2ll} and \ref{eq:V1V2llll}, which in the case of leptonic decays remains valid in the presence of higher-order QCD corrections, relies on the actual presence of an EW boson which is produced in $s$-channel and then decays into two leptons. Diagrams characterised by this structure are often dubbed \emph{resonant}. Strictly speaking, such a situation is physical only if intermediate EW boson is on-shell \cite{Denner:2020bcz}. 
Resonant and non-resonant diagrams contributing to four-lepton production at the LHC are shown in Fig.~\ref{fig:ZZdiagrams}.
While in general given a certain final state, \emph{e.g.} four charged leptons, both resonant and non-resonant diagrams contribute to the full off-shell process, very often the resonant ones give the dominant contribution.
This is the case for inclusive di-boson production for which full off-shell results differ from those computed in the pole \cite{Denner:2020bcz,Denner:2020eck,Poncelet:2021jmj,Denner:2021csi,Denner:2023ehn} or in the narrow-width approximation \cite{Poncelet:2021jmj,Hoppe:2023uux} by less than 2\% in most of phase-space regions.[, as expected from the intrinsic accuracy of the employed on-shell approximations \cite{Denner:2000bj,Uhlemann:2008pm}.]

If the resonant topologies and therefore the on-shell-boson phase-space regions are dominant, one expects that the angular coefficients extracted from full off-shell distributions according to Eqs.~\ref{eq:V2ll} and \ref{eq:V1V2llll} are very close to those found for on-shell calculations, as long as two-body decays are present.
This is confirmed by the numerical results in Table
\ref{tab:coeff_coscos}, where we show a comparison of selected angular coefficients obtained with full off-shell and double-pole-approximation (DPA) simulations. 
\begin{table}[t]
  \begin{center}
\begin{tabular}{crrrr}
\hline
\cellcolor{green!9} & \multicolumn{2}{c}{\cellcolor{green!9} ZZ} & \multicolumn{2}{c}{\cellcolor{green!9} W$^+$Z} \\
\hline\rule{0ex}{2.7ex}
       \cellcolor{blue!9}  
       & \cellcolor{blue!9} full off-shell
       & \cellcolor{blue!9} DPA
       & \cellcolor{blue!9} full off-shell
       & \cellcolor{blue!9} DPA   \\
       \hline\\[-0.3cm]
$\alpha^{(1)}_{1,0}$	 	 &  $-0.00093 (5 )$  & $  -0.00098 (5 )$ & $-0.0464 (3 )$ & $-0.0472 (3 )$\\[0.1cm]
$\alpha^{(1)}_{2,0}$		 &  $0.02807 (5 )$ 	 & $  0.02806 (5 )$ & $0.0283 (3 )$  & $0.0289 (3 )$\\[0.1cm]
$\alpha^{(3)}_{1,0}$ 	 &  $-0.00102 (5 )$  & $  -0.00109 (5 )$ & $0.0040 (3 )$ & $0.0041 (3 )$\\[0.1cm]
$\alpha^{(3)}_{2,0}$	 	 &  $0.02794 (5 )$ 	 & $  0.02802 (5 )$ & $0.0298 (3 )$ & $0.0294 (3 )$ \\[0.1cm]
$\gamma_{1,0,1,0 }$ 	 &  $-0.00146 (2 )$  & $  -0.00150 (2 )$ & $-0.0052 (1 )$ & $-0.0053 (1 )$\\[0.1cm]
$\gamma_{2,0,2,0 }$ 	 &  $0.00167 (1 )$ 	 & $  0.00168 (2 )$ & $0.0012 (1 )$  & $0.0014 (1 )$  \\[0.1cm]
$\alpha^{(1)}_{2,-2}$	 	 &  $-0.01000 (2 )$  & $  -0.01002 (2 )$ & $-0.0120 (3 )$ &$-0.0118 (3 )$ \\[0.1cm]
$\alpha^{(3)}_{2,-2}$ 	 &  $-0.01001 (2 )$  & $  -0.00999 (2 )$ & $-0.0074 (3 )$ & $-0.0078 (4 )$ \\[0.1cm]
\hline
\end{tabular}  
    \end{center}
  \caption{
  Angular coefficients for $\PZ\PZ$ and $\PW^+\PZ$ at NLO QCD accuracy in the full off-shell and DPA calculation, computed according to Eq.\ref{eq:inc}. The label $1(3)$ is associated to coefficients associated to the first (second) boson.
  The inclusive setup defined in Eq.~\eqref{eq:inclsetup} is understood. Numerical-integration uncertainties are shown in parentheses.
  A fixed renormalisation and factorisation scale $\mu_{\rm F}=\mu_{\rm R}=(M_{V_1}+M_{V_2})/2$ is assumed. \label{tab:coeff_coscos}}
\end{table}
Both $\PZ\PZ$ and $\PW^+\PZ$ channels are considered.
It is striking to see how all coefficients computed in the DPA agree within Monte Carlo uncertainties with those computed including off-shell effects.
It is worth noticing that forcing same-flavour, opposite-sign leptons to have an invariant-mass sufficiently close to the $\PZ$-boson mass further suppresses non-resonant contributions, making the 
DPA results even closer to the off-shell ones.
If such constraints are not applicable, as in the $\PW$-boson leptonic decay, slightly larger off-shell effects have to be expected in some phase-space regions \cite{Biedermann:2016guo,Biedermann:2017oae,Denner:2020bcz,Denner:2020eck}. In sufficiently inclusive setups, even for $\PW$ bosons the DPA and the full off-shell one give compatible results, as shown by the coefficients $\alpha^{(1)}_{1,0}$ and $\alpha^{(1)}_{2,0}$ in the two rightmost columns of Table~\ref{tab:coeff_coscos}. 

The results shown are NLO accurate in QCD,
and have been computed integrating fully differential weights from {\scshape POWHEG-BOX-RES} \cite{Pelliccioli:2023zpd} multiplied by suitable combinations of spherical harmonics ($\ell\leq 2$) in the decay angles, according to Eq.~\ref{eq:inc}.

As we are going to analyse in Sect.~\ref{sec:FOrad} 
the QCD corrections do not modify the analytic structure of Eq.~\ref{eq:V1V2llll} in the case of leptonic decay, while NLO EW corrections do, introducing non-factorisable corrections which are by definition only part of the full off-shell calculation. However, it has been shown that the DPA description of di-boson processes 
reproduces full off-shell results by order 1\%
also at NLO EW, in both inclusive and fiducial setups \cite{Denner:2021csi,Le:2022lrp,Le:2022ppa,Denner:2023ehn,Dao:2023pkl}.
As a numerical check of this expectation we have 
calculated the coefficient $\alpha^{(1)}_{2,0}$ at NLO EW both in the DPA and in the full off-shell case with the techniques recently developed in the MoCaNLO framework \cite{Denner:2021csi,Denner:2023ehn}. For comparison, in Table~\ref{tab:coeff_a_1_20} we report these NLO EW results together with LO ones and the NLO QCD ones already shown in Table~\ref{tab:coeff_coscos}.
\begin{table}[t]
  \begin{center}
\begin{tabular}{lllllll}
\hline
\cellcolor{green!9} 
& \multicolumn{3}{c}{\cellcolor{green!9} $\sigma$} 
& \multicolumn{3}{c}{\cellcolor{green!9} $\alpha^{(1)}_{2,0}$} \\
\hline\rule{0ex}{2.7ex}
       \cellcolor{blue!9}  
       & \cellcolor{blue!9} full off-shell
       & \cellcolor{blue!9} DPA   
       & \cellcolor{blue!9} $\delta$   
       & \cellcolor{blue!9} full off-shell
       & \cellcolor{blue!9} DPA   
       & \cellcolor{blue!9} $\delta$\\
       \hline\\[-0.3cm]
LO	 & $ 21.483(8)$ & $ 21.209(7)$ & $-1.28\%$ & $0.02977 ( 2  )$ & $ 0.02982 ( 3 )$ & $+0.16\%$ \\[0.1cm]
NLO QCD	  & $28.63(1)$ & $28.22(1)$ & $-1.43\%$ & $0.02807 (5 )$ & $  0.02806 (5 )$ & $-0.04\%$ \\[0.1cm]
NLO EW	 & $19.15(1)$& $18.94(1)$& $-1.10\%$ & $0.02932 (5)$ & $ 0.02944 ( 5 ) $ & $ +0.41\%$\\[0.1cm]
\hline
\end{tabular}  
    \end{center}
  \caption{
  Integrated cross section ($\sigma_{\rm inc}$) and coefficient $\alpha^{(1)}_{2,0}$ for $\PZ\PZ$ in the full off-shell and DPA calculation. 
  $\delta$ (shown in percentage) is defined as the difference between the DPA and the full off-shell result, normalised to the full off-shell one.
  Results have been computed with {\scshape MoCaNLO} \cite{Denner:2021csi,Denner:2023ehn}.  The inclusive setup defined in Eq.~\eqref{eq:inclsetup} is understood. Numerical uncertainties are shown in parentheses.
  A fixed renormalisation and factorisation scale $\mu_{\rm F}=\mu_{\rm R}=\MZ$ is assumed. \label{tab:coeff_a_1_20}}
\end{table}
One can see how the discrepancy between the DPA and full off-shell results found for of total cross sections is at the $1\%$ level, while the analogous discrepancy for the $\alpha^{(1)}_{2,0}$ is at the sub-percent level, with the largest one ($0.4\%$) found at NLO EW. 

We have proven that the off-shell description of inclusive di-boson production and decay is well reproduced by the DPA,
also for the radiative corrections, whose impact is discussed in Sect.~\ref{sec:FOrad}.

A word of caution is needed for the decay of a SM Higgs boson into four leptons. Owing to a $125$-GeV pole mass, it is not possible to have two intermediate $\PZ$ bosons simultaneously on-shell. At tree level this is solved \cite{Aguilar-Saavedra:2022wam,Barr:2022wyq,Fabbrichesi:2023cev} assigning a fictitious mass, lower than $\MZ$, to the off-shell boson, which anyway shows up in $s$-channel with the same spherical-harmonic structure of Eq.~\ref{eq:V1V2llll}. However, this assumption is not well defined anymore after including finite-width and NLO EW effects. A detailed analysis of these aspects is presented in Sect.~\ref{sec:HiggsHO}.

\subsection{Higher-order corrections}\label{sec:FOrad}
In Sect.~\ref{sec:offshell} we have shown that the complete description of four leptons production including non-resonant contributions gives the same results as on-shell approximations for angular coefficients extracted according to the angular structure of Eq.~\ref{eq:V1V2llll}, both at LO and with radiative corrections. However, we consider in the following the complete matrix elements for four-lepton production and decay, to systematically include subleading effects, which might be relevant in the precision determination of entanglement parameters.

Including higher-order corrections is crucial for the correct interpretation of SM predictions with LHC data.
For the $\PZ\PZ$ process both QCD and EW corrections are rather large and therefore crucial for the proper modeling of both total cross section and distribution shapes. In the inclusive setup (see Eq.~\ref{eq:inclsetup}) the relative NLO corrections read, for the full off-shell description, 
\beq
\delta^{\rm NLO}_{\rm EW}=-11\% \,,\quad \delta^{\rm NLO}_{\rm QCD}=+33\%\,.
\eeq
Very similar values are found in the fiducial setup of Eq.~\ref{eq:ATLASfid}.
It is worth recalling that the $\PZ\PZ$ receives a contribution from the $\Pg\Pg$ channel which, though formally of NNLO accuracy in QCD, gives a $+15\%$ enhancement to the integrated cross section. The complete NNLO QCD corrections are known \cite{Grazzini:2015hta, Cascioli:2014yka} but not considered in our work.

In the considered case of leptonic decays, the QCD corrections only come from initial-state radiation (ISR) as shown in the sample diagrams in Fig.~\ref{fig:NLOQCDdiag}.
It is in fact well known \cite{Bern:2011ie} that without cuts applied to the leptons the angular structure in Eqs.~\ref{eq:V2ll} and \ref{eq:V1V2llll} remains valid after including QCD corrections, because it depends only on the interaction vertex of the $Z$ boson with the leptons. In other words, the formal structure of the spin-density matrix does not change between LO and (N)NLO QCD. 
However, the QCD corrections change the production rate in the leading $q\bar{q}$ contribution and open up new gluon-induced partonic channels. Effectively, this translates into a different relative weight of the various helicity contributions to the $\PZ\PZ$ production cross section compared to LO. Therefore, a change is to be expected in the value of angular coefficients.
This can be clearly seen comparing the LO and NLO QCD numerical results shown in Table~\ref{tab:coeff_nloew_zz}
for a small set of coefficients.
\begin{table}
  \begin{center}
    \begin{tabular}{crrr}
    \hline
    \cellcolor{green!9} & \multicolumn{3}{c}{\cellcolor{green!9} ZZ} \\
      \hline\rule{0ex}{2.7ex}
       \cellcolor{blue!9}  
       & \cellcolor{blue!9} LO
       & \cellcolor{blue!9} NLO QCD   
       & \cellcolor{blue!9} NLO EW \\
       \hline\\[-0.3cm]
$\alpha^{(1)}_{1,0}$ 	 &  $-0.00001 (9 )$ 	 &  $-0.00097 (10 )$ 	 & $  -0.00004 (9 )$ \\[0.1cm]
$\alpha^{(1)}_{2,0}$ 	 &  $0.03009 (11 )$ 	 &  $0.02794 (13 )$ 	 & $  0.02960 (11 )$ \\[0.1cm]
$\alpha^{(3)}_{1,0}$ 	 &  $0.00012 (13 )$ 	 &  $-0.00086 (15 )$ 	 & $  0.00018 (14 )$ \\[0.1cm]
$\alpha^{(3)}_{2,0}$ 	 &  $0.03006 (7 )$ 	 &  $0.02796 (6 )$ 	 & $  0.02964 (7 )$ \\[0.1cm]
$\gamma_{1,0,1,0 }$ 	 &  $-0.00173 (3 )$ 	 &  $-0.00148 (3 )$ 	 & $  -0.00043 (3 )$ \\[0.1cm]
$\gamma_{2,0,2,0 }$ 	 &  $0.00188 (2 )$ 	 &  $0.00168 (2 )$ 	 & $  0.00187 (2 )$  \\[0.1cm]
$\alpha^{(1)}_{2,-2}$  	 &  $-0.00967 (7 )$ 	 &  $-0.00993 (9 )$ 	 & $  -0.00991 (7 )$  \\[0.1cm]
$\alpha^{(3)}_{2,-2}$  	 &  $-0.00973 (4 )$ 	 &  $-0.01003 (4 )$ 	 & $  -0.00996 (4 )$  \\[0.1cm]
\hline
    \end{tabular}  
    \end{center}
  \caption{
  Angular coefficients for $\PZ\PZ$ at LO, NLO QCD and NLO EW accuracy in the full off-shell process. All numbers have been computed integrating fully differential weights from {\scshape POWHEG-BOX-RES} \cite{Chiesa:2020ttl} multiplied by suitable combinations of spherical harmonics ($\ell\leq 2$) in the decay angles, according to Eq.~\ref{eq:coscos}.  The inclusive setup defined in Eq.~\eqref{eq:inclsetup} is understood. Numerical uncertainties are shown in parenthesis.
  A fixed renormalisation and factorisation scale $\mu_{\rm F}=\mu_{\rm R}=\MZ$ is assumed.
  \label{tab:coeff_nloew_zz}}
\end{table}
The single-boson polar coefficients $\alpha_{2,0}$ are diminished by 7\%, owing to the opening of the gluon-induced channel which increase the fraction of events with at least one longitudinal boson \cite{Denner:2021csi}.
For the same motivation, the almost perfect balance between left and right helicity of each boson found at LO is not there at NLO QCD \cite{Denner:2021csi}, resulting in small but not vanishing $\alpha_{1,0}$ coefficients.
The polar spin-correlation coefficients $\gamma_{2,0,2,0}$ and $\gamma_{1,0,1,0}$ are 
strongly affected by QCD radiative corrections, with $10$-to-$15\%$ changes. 
Notice that the correlation coefficient $\gamma_{1,0,1,0}$ is finite and negative although the $\alpha_{1,0}$'s are either compatible with zero (LO) or suppressed (NLO QCD). This is motivated by the left-right (and right-left) transverse modes of the bosons which dominate in the $q\bar{q}$ channel in the SM ($f_{\pm\pm} \approx 2\% \ll f_{\pm\mp} \approx 35\%$). 
The QCD effects are less severe but not negligible ($3\%$) for the considered CP-even azimuthal coefficients $\alpha_{2,-2}$.
The inclusion of the QCD corrections is essential when measuring spin-correlation probes like the $R_c$ pseudo-observable defined in Eq.~\ref{eq:Rcdef}. Their inclusion sizeably diminishes the $R_c$ value for $\PZ\PZ$,
\beq
R_c^{\rm \,(LO)} \,=\, 1.89(3)\qquad
R_c^{\rm \,(NLO \,QCD)} \,=\,1.73(2)\,,
\eeq 
where numerical uncertainties come from the standard error propagation from coefficients in Table~\ref{tab:coeff_nloew_zz}.
It is worth noticing that the effect of QCD corrections on $R_c$ is even stronger for the $\PW^+\PZ$ process, 
\beq
R_c^{\rm \,(LO)} \,=\, 2.39(3)\qquad
R_c^{\rm \,(NLO \,QCD)} \,=\,1.31(2)\,.
\eeq 
We remark that this sizeable effect coming from NLO QCD corrections is not covered by the
LO-uncertainty estimate based on QCD-scale and PDF variations.

The inclusion of EW radiative corrections in the modeling of angular coefficients makes the picture a bit more involved than the one found at (N)NLO in QCD. 
In the presence of additional photons coming from real corrections, it is in fact crucial to define the decay angles as infrared-safe quantities.
We choose to consider the kinematics of dressed leptons,  reconstructed through a cone dressing with $R_{\ell\gamma}=0.1$ resolution radius. This strategy is well defined theoretically and already used in ATLAS and CMS analyses. Real photons can be emitted both from initial-state quarks and from final-state leptons, as depicted in Fig.~\ref{fig:NLOQEDLOdiag}, thus changing both the production and the decay structure of the di-boson process.
In particular, real photons emitted from the final state (FSR) change the analytic structure of Eq.~\ref{eq:V2ll}, making the boson undergo a three-body decay and therefore introducing higher-rank spherical harmonics.
Also one-loop EW corrections affect both the production and the decay of EW bosons. They introduce new non-resonant topologies absent at LO, up to 5- and 6-point functions (second and third diagram in Fig.~\ref{fig:NLOEWdiag}), but also box corrections to the production sub-amplitude (analogous to the QCD ones, first diagram in Fig.~\ref{fig:NLOEWdiag}), and triangle corrections to the two-body decays of weak-boson decays (fourth diagram in Fig.~\ref{fig:NLOEWdiag}).
This means that in principle larger effects are expected than at NLO QCD, but suppressed due to power counting.
The results in Table~\ref{tab:coeff_nloew_zz}, obtained with full off-shell calculations, show that the effect of EW corrections on single-boson polar coefficients is small ($1.5$\% on $\alpha_{2,0}$), slightly larger for azimuthal coefficients ($2.5$\% on $\alpha_{2,0}$). While negligible effects are found for $\gamma_{2,0,2,0}$, the EW corrections completely change the LO transverse-helicity balance, with a marked decrease (in absolute value) for $\gamma_{1,0,1,0}$.

We have proven that the inclusion of the NLO EW corrections is unavoidable for a sound description of angular coefficients.
More specifically, it has been shown \cite{Denner:2021csi} that the dominant effects on decay-angle
differential distributions come from one-loop factorisable corrections and ISR real radiation.
The QED FSR effects on angular coefficients are minor, as far as dressed leptons are concerned.

One last remark is in order. Although we have only discussed fixed-order corrections to
the tree-level picture of di-boson processes. The parton-shower effects have been found to be negligible for the extraction of angular coefficients. This has been numerically checked by matching the NLO calculations of the $\PZ\PZ$ and $\PW\PZ$ processes 
to the QCD+QED {\scshape Pythia}8 shower \cite{Sjostrand:2019zhc} within the {\scshape PowHeg-Box} di-boson package \cite{Chiesa:2020ttl,Pelliccioli:2023zpd}. 

\subsection{Higgs-boson decay}\label{sec:HiggsHO}
In this section we scrutinise the angular coefficients in the four-lepton decay of a SM Higgs boson, for which the discussion of off-shell effects and the one of NLO EW corrections are strictly connected.

\begin{figure}
  \centering
   \subfigure[LO and NLO EW (real)\label{fig:LOdiagH}]{\includegraphics[scale=0.35]{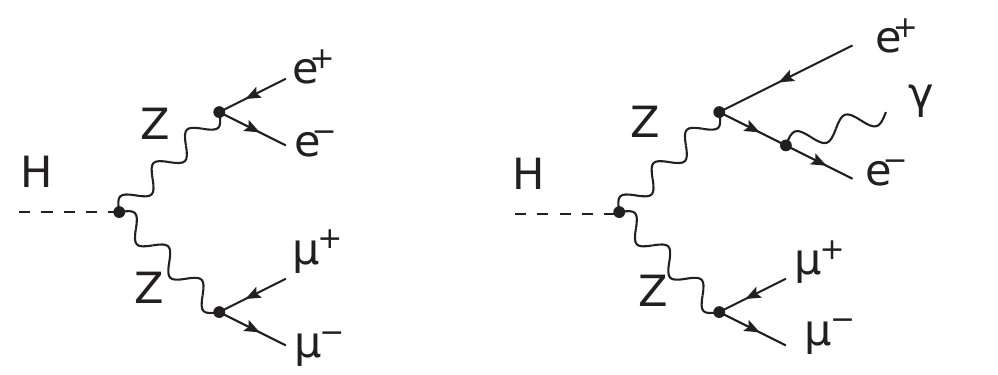}}
   \subfigure[NLO EW (virtual)\label{fig:NLOEWdiagH}]{\includegraphics[scale=0.35]{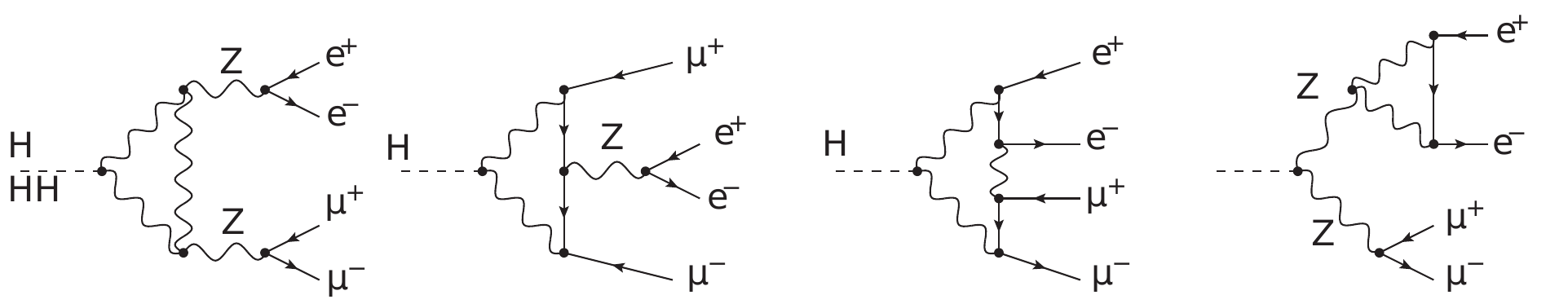}}\\
  \caption{Sample diagrams contributing to the Higgs-boson decay into four charged leptons. A tree-level and diagram is shown (a, left), along with a real-radiation (a, right) and one-loop diagrams (b).\label{fig:Hdiagrams}}
\end{figure}

Due to the very narrow width, we ignore interference and off-shell effects associated to the Higgs-boson modeling and focus on its on-shell regime. The decay process is shown in Eq.~\ref{eq:Higgsdec} and sample diagrams are depicted in Fig.~\ref{fig:Hdiagrams}.

Owing to a pole mass of $125\GeV$, an on-shell Higgs boson cannot decay into two on-shell $\PZ$ bosons.
Thanks to the simple SM tree-level amplitude structure with two $s$-channel $\PZ$-boson propagators (left diagram in Fig.~\ref{fig:LOdiagH}), the angular expansion of Eq.~\ref{eq:V1V2llll} is still valid \cite{Aguilar-Saavedra:2022wam,Fabbrichesi:2023cev}.
This is no longer true at NLO EW, due to the appearance of non-factorisable corrections, like the second and third diagram in Fig.~\ref{fig:NLOEWdiagH}. In fact, the factorisable corrections are expected to be dominant for the on-shell $\PZ$ boson, leading to a reliable estimate of the coefficients associated to it. This can be proven explicitly through a single-pole approximation \cite{Maina:2021xpe,Chen:2021sgd}. On the contrary, it is not guaranteed that the factorisable corrections are dominant if the lepton pair is off-shell. 
Therefore, since Eq.~\ref{eq:V1V2llll} assumes two intermediate $\PZ$ bosons in $s$-channel both dominated by on-shell kinematics, at NLO EW the analytical structure of Eq.~\ref{eq:V1V2llll} is no longer valid, owing to missing higher-rank contributions.
If we stick to LO structure, we implicitly embed higher-rank effects into $l\leq 2$ coefficients, thus obtaining large modifications of their numerical value. 
This suggests that the interpretation of $l\leq 2$ coefficients as if they were describing a system of two spin-1 massive bosons is misleading. In turn, the translation into quantum observables may be significantly affected.

As introduced for inclusive di-boson production in Sect.~\ref{sec:FOrad}, the lepton dressing with real photons is needed for a theoretically sound, \emph{i.e.} infrared-safe, decay-angle definition.  
Additionally, in the Higgs-boson decay (but also in general) the presence of real photons at NLO EW (or from QED parton shower) opens up two possibilities for the reference frame where the spin of intermediate $\PZ$ bosons is quantised: (i) the Higgs-boson rest frame (H-rest), and (ii) the four-lepton CM frame (4$\ell$-CM). The two frames coincide at LO, but differ at NLO EW by real contributions with a photon that is not recombined with charged leptons. While the former option seems more natural from a theoretical viewpoint, the latter is more realistic in a collider environment, \emph{i.e.} when the Higgs boson is boosted.

According to the simple tree-level structure of the CP-conserving SM amplitude with massless leptons, \emph{i.e.} first diagram in Fig.~\ref{fig:LOdiagH}, and to our definition of decay angles (see Fig.~\ref{fig:geomscheme}), just a few angular coefficients do not vanish \cite{Aguilar-Saavedra:2022wam} and they are related to each other through the relations, \beqn\label{eq:symmHiggs1}
&\alpha_{2,0}^{(1)} \,=\, \alpha_{2,0}^{(3)}\,,\quad \gamma_{2,2,2,-2} \,=\, \gamma_{2,-2,2,2}\,,
\quad\gamma_{2,1,2,-1} \,=\, \gamma_{2,-1,2,1}\,, \quad\gamma_{1,1,1,-1} \,=\, \gamma_{1,-1,1,1}\,,&\\
& 20\pi\left(\gamma_{2,2,2,-2}+\gamma_{2,0,2,0}\right) \,=\, 1 
\,, \qquad 
40\pi\gamma_{2,2,2,-2} \,=\, \sqrt{20\pi}\alpha_{2,0}^{(1)} + 1
\,,& \label{eq:symmHiggs2}\\
&5\eta_\ell^2\,\gamma_{2,1,2,-1} \,=\, \gamma_{1,1,1,-1}
\,,\qquad
\eta_\ell^2\,\left(1-20\pi\,\gamma_{2,0,2,0}\right) \,=\, {4\pi}\gamma_{1,0,1,0}
\,. &  \label{eq:symmHiggs3}
\eeqn
In Table~\ref{tab:coeff_nloew_zzH} we present the angular coefficients obtained at LO and NLO EW with the 4$\ell$-CM and H-rest frame choices. In addition, numerical results are shown in both the default setup defined in Eq.~\ref{eq:mll10} and in a setup where the electron-positron pair is in the on-shell region,
\beq\label{eq:asymmsetupH}
81\GeV<M_{\Pe^+\Pe^-}<101\GeV\,,\qquad M_{\mu^+\mu^-}> 10\GeV\,.
\eeq
\begin{table}[t]
  \begin{center}
\begin{tabular}{lrrr}
\hline
\cellcolor{green!9} &\cellcolor{green!9} LO &\cellcolor{green!9} NLO EW (4$\ell$-CM) &\cellcolor{green!9} NLO EW (H-rest)     \\
\hline
\cellcolor{blue!9} &     \multicolumn{3}{c}{\cellcolor{blue!9} $M_{\Pe^+\Pe^-}>10\GeV$, $M_{\mu^+\mu^-}>10\GeV$}\\
\hline\rule{0ex}{2.7ex}\\[-0.4cm]
$\alpha^{(1)}_{2,0}$ &  $-0.04792 (4 )$ &  $-0.04498 (6 )$ &  $-0.04215 (6 )$ \\[0.1cm]
$\alpha^{(3)}_{2,0}$ &  $-0.04792 (4 )$ &  $-0.04506 (6 )$ &  $-0.04224 (6 )$ \\[0.1cm]
$\gamma_{1,0,1,0 }$ 	 &  $0.00117 (1 )$ &  $0.00012 (2 )$ &  $0.00011 (2 )$ \\[0.1cm]
$\gamma_{2,0,2,0 }$ 	 &  $0.01097 (1 )$ &  $0.01079 (2 )$ &  $0.01019 (2 )$ \\[0.1cm]
$\gamma_{1,1,1,-1 }$ 	 &  $-0.00185 (2 )$ &  $-0.00048 (2 )$ &  $-0.00047 (2 )$ \\[0.1cm]
$\gamma_{1,-1,1,1 }$ 	 &  $-0.00186 (2 )$ &  $-0.00047 (2 )$ &  $-0.00047 (2 )$ \\[0.1cm]
$\gamma_{2,1,2,-1 }$ 	 &  $-0.00776 (1 )$ &  $-0.00779 (2 )$ &  $-0.00715 (2 )$ \\[0.1cm]
$\gamma_{2,-1,2,1 }$ 	 &  $-0.00778 (1 )$ &  $-0.00783 (2 )$ &  $-0.00709 (2 )$ \\[0.1cm]
$\gamma_{2,2,2,-2 }$ 	 &  $0.00493 (2 )$ &  $0.00489 (2 )$ &  $0.00481 (2 )$ \\[0.1cm]
$\gamma_{2,-2,2,2 }$ 	 &  $0.00494 (2 )$ &  $0.00488 (2 )$ &  $0.00479 (2 )$ \\[0.1cm]
\hline
\cellcolor{blue!9}&\multicolumn{3}{c}{\cellcolor{blue!9} $81\GeV<M_{\Pe^+\Pe^-}<101\GeV$, $M_{\mu^+\mu^-}>10\GeV$}\\
\hline\rule{0ex}{2.7ex}\\[-0.4cm]
$\alpha^{(1)}_{2,0}$ &  $-0.04395 (8 )$ &  $-0.0435 (1 )$ &  $-0.0430 (1 )$ \\[0.1cm]
$\alpha^{(3)}_{2,0}$ &  $-0.04385 (8 )$ &  $-0.0418 (1 )$ &  $-0.0412 (1 )$ \\[0.1cm]
$\gamma_{1,0,1,0 }$ 	 &  $0.00122 (2 )$ &  $-0.00023 (3 )$ &  $-0.00024 (3 )$ \\[0.1cm]
$\gamma_{2,0,2,0 }$ 	 &  $0.01071 (2 )$ &  $0.01064 (3 )$ &  $0.01045 (3 )$ \\[0.1cm]
$\gamma_{1,1,1,-1 }$ 	 &  $-0.00189 (3 )$ &  $0.00012 (5 )$ &  $0.00012 (5 )$ \\[0.1cm]
$\gamma_{1,-1,1,1 }$ 	 &  $-0.00189 (3 )$ &  $0.00017 (5 )$ &  $0.00016 (5 )$ \\[0.1cm]
$\gamma_{2,1,2,-1 }$ 	 &  $-0.00794 (3 )$ &  $-0.00798 (4 )$ &  $-0.00773 (4 )$ \\[0.1cm]
$\gamma_{2,-1,2,1 }$ 	 &  $-0.00797 (3 )$ &  $-0.00801 (4 )$ &  $-0.00771 (4 )$ \\[0.1cm]
$\gamma_{2,2,2,-2 }$ 	 &  $0.00518 (3 )$ &  $0.00515 (5 )$ &  $0.00511 (5 )$ \\[0.1cm]
$\gamma_{2,-2,2,2 }$ 	 &  $0.00521 (3 )$ &  $0.00516 (5 )$ &  $0.00509 (5 )$ \\[0.1cm]
\hline
\end{tabular}  
    \end{center}
  \caption{
 Angular coefficients for a SM Higgs-boson decay into four charged leptons (see Eq.~\eqref{eq:Higgsdec}) at LO and NLO EW accuracy in the default setup of Eq.~\eqref{eq:mll10} and the one of Eq.~\ref{eq:asymmsetupH}. Numerical uncertainties are shown in parenthesis. For NLO EW results two frames are chosen for the quantisation axis of the intermediate-boson spin: the CM frame of the four dressed leptons ($4\ell$-CM) and the Higgs rest frame (H-rest).
  \label{tab:coeff_nloew_zzH}}
\end{table}
The relations of Eq.~\ref{eq:symmHiggs1} are confirmed numerically at LO in both setups. In the default setup, for which the invariant mass of both fermion pairs is integrated over the whole available range ($10\GeV< M_{\ell^+\ell^-}<\sqrt{125^2-10^2}\GeV$), the equalities are fulfilled also at NLO EW, with both definitions of the angular variables (4$\ell$-CM and H-rest). In the setup of Eq.~\ref{eq:asymmsetupH}, the $\alpha_{2,0}$ coefficients are different, with $\alpha^{(1)}_{2,0}$, associated to the boson in the on-shell region, getting closer to the LO value than $\alpha^{(3)}_{2,0}$. This effect comes from one-loop contributions with no $s$-channel $\PZ$-boson propagator associated to the $\PZ^*\rightarrow \mu^+\mu^-$ off-shell decay. 

The first tree-level relation of Eq.~\ref{eq:symmHiggs2} is mildly violated by NLO EW corrections especially if decay angles are defined in the H-rest frame and the default setup is considered. The 4$\ell$-CM definition gives more LO-like results in both setups. The second relation of Eq.~\ref{eq:symmHiggs2} is violated at NLO EW for both frame choices.

Both equalities of Eq.~\ref{eq:symmHiggs3} involve $(l=2,m)$ coefficients on the left side and the corresponding $(l=1,m)$ coefficients on the right side. A mild modification of $(l=2,m)$ coefficients is found at NLO EW, while strikingly large EW corrections characterise the $(l=1,m)$ coefficients. Such a large change, and in particular the sign flip found in the setup of Eq.~\ref{eq:asymmsetupH}, cannot come from one-loop corrections to $\sin^2\theta^{\rm \,eff}_\ell$ \cite{Degrassi:1996ps} and their implication on $\eta_\ell$, therefore they can be traced back to large non-factorisable effects that clearly break the tree-level angular structure.

In Sect.~\ref{sec:AngCoeff} we have highlighted the role of the reference-frame choice in the
angular-coefficient extraction. The results of Table~\ref{tab:coeff_nloew_zzH}, especially for what concerns $l=2$ coefficients illustrates the non-negligible impact of different frame choice at NLO EW. The radiative effects are the 6.5\% level for $\alpha_{2,0}$ in the 4$\ell$-CM frame, while 14\% corrections are found in the H-rest frame. Similarly, while $\gamma_{2,0,2,0}$ receives a $-1.5\%$ correction in the 4$\ell$-CM frame, $-7\%$ is found in the H-rest one. Also for azimuthal-dependent coefficients the NLO EW values are sizeably closer to the LO ones if the 4$\ell$-CM frame is considered. In the setup of Eq.~\ref{eq:asymmsetupH} the differences between the two frame choices decrease to the $1\%$ level, although the results in the 4$\ell$-CM frame are again closer to the LO counterparts.

As discussed in Sect.~\ref{sec:quantumobs}, two spin-1 bosons produced in the decay of a scalar particle, are in an entangled state if and only if $\gamma_{2,1,2,-1} \neq 0$ or $\gamma_{2,2,2,-2} \neq 0$ \cite{Aguilar-Saavedra:2022wam}. From Table~\ref{tab:coeff_nloew_zzH}, it can be seen that the entanglement level is sizeable. Additionally, the NLO EW value for the two azimuthal coefficients are numerically compatible with the corresponding LO one as far as the 4$\ell$-CM frame is concerned. For the $\gamma_{2,\pm 1, 2, \mp 1}$ coefficients the H-rest frame choice leads to NLO EW corrections of about $-8\%$ and $-4\%$ for the two considered setups. 
While the LO results are likely to be enough to claim entanglement in the Higgs-boson decay, according to the condition in Eq.~\ref{eq:entangcond}, for a precise evaluation of the level of entanglement it is unavoidable to include EW corrections and to define carefully the reference frame for the spin quantisation.
For what concerns Bell-inequality violation, we have evaluated the quantity $I_3$, defined in Eq.~\ref{eq:I3explicit}, at LO and at NLO EW in different reference frames, obtaining
\beqn
{\rm LO}:                     && I_3 \,=\, 2.671(2)\,,\nnb\\
\textrm {NLO\,EW, 4$\ell$-CM}:&& I_3 \,=\, 2.682(4)\,,\\
\textrm {NLO\,EW, H-rest}:    && I_3 \,=\, 2.571(4)\,.\nnb
\eeqn
The LO result is hardly changed by EW corrections in the 4$\ell$-CM frame, slightly larger effects ($-4\%$) are found in the H-rest frame. Similarly to the case of the entanglement conditions, assuming the tree-level expansion of the spin-density matrix and using the LO results is likely to be sufficient to claim the violation of the CGLMP Bell inequality. 

\begin{figure}[t]
  \centering
    \includegraphics[scale=0.36]{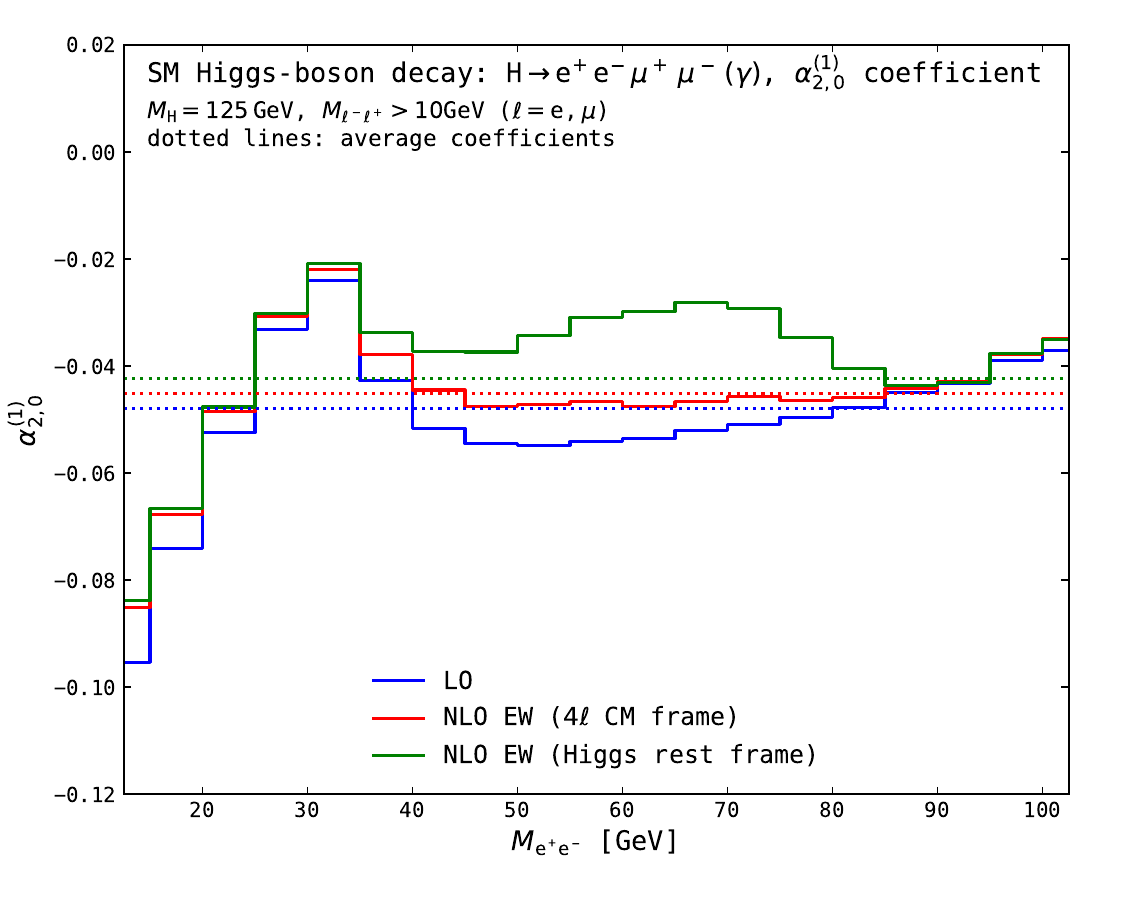}   
    \includegraphics[scale=0.36]{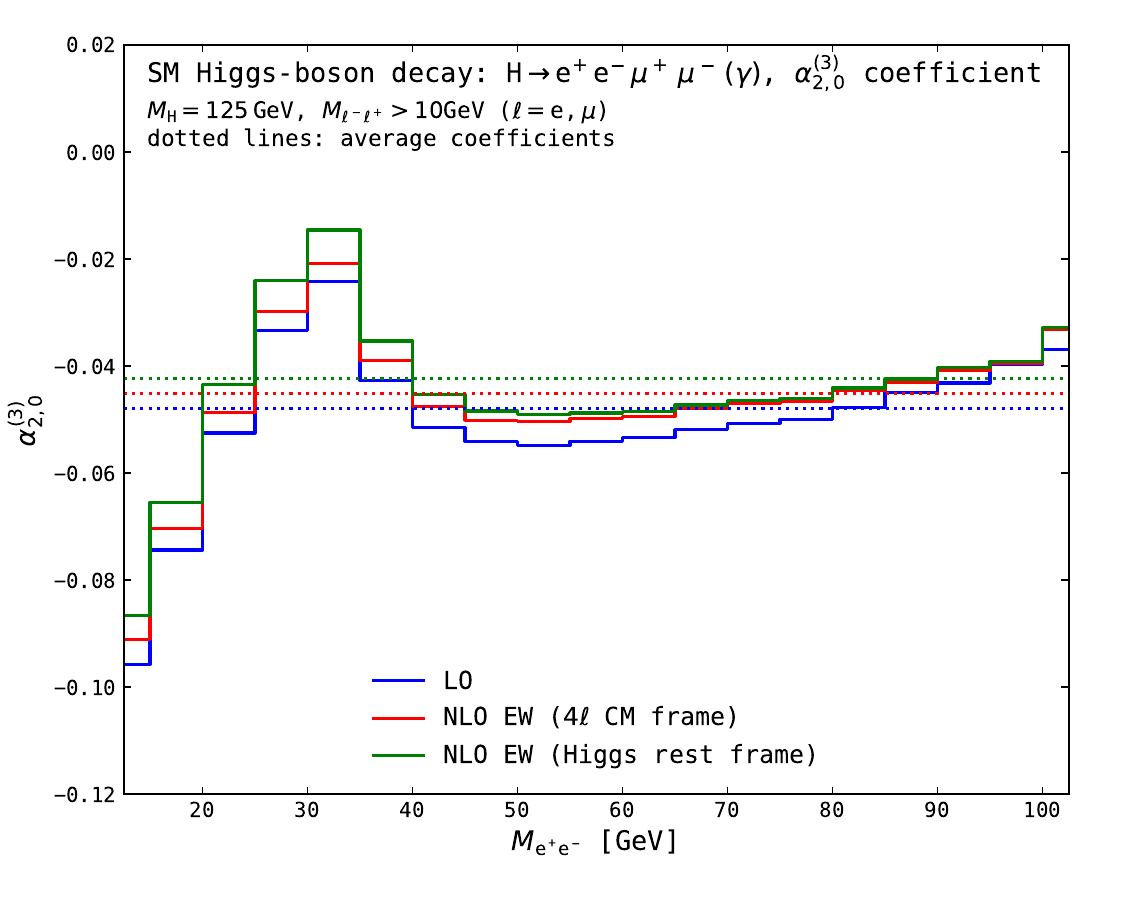}
  \caption{
  Dependence of the polar coefficients $\alpha_{2,0}^{(1)}$ (left plot, associated to $\PZ^{(*)}\!\rightarrow \Pe^+\Pe^-$) and $\alpha_{2,0}^{(3)}$ (right plot, associated to $\PZ^{(*)}\!\rightarrow \mu^+\mu^-$) on the off-shell-ness of the $\Pe^+\Pe^-$ pair ($M_{\Pe^+\Pe^-}$), in the Higgs-boson decay into four charged leptons (see Eq.~\eqref{eq:Higgsdec}) at LO and NLO EW accuracy. In the case of NLO EW results, two different choices are made for the reference frame where polarisations of intermediate $\PZ^{(*)}$ bosons are defined: the four-lepton CM frame (red) and the Higgs-boson rest frame (green).
  Dashed lines are reference values obtained integrating over $M_{\Pe^+\Pe^-}$.
  }\label{fig:MassDependence}
\end{figure}

As a last comment of this section, we further scrutinise the relation between the off-shell-ness of same-flavour lepton pairs and the corresponding $\alpha_{2,0}$  coefficients.
In Fig.~\ref{fig:MassDependence} we consider the values of $\alpha_{2,0}^{(1)}$ (left, associated to the $\Pe^+\Pe^-$ pair) and $\alpha^{(3)}_{2,0}$ (right, associated to the $\mu^+\mu^-$ pair), as functions of the invariant mass of the $\Pe^+\Pe^-$ system.
For reference, we show as dashed flat lines the results obtained in the default setup (first two rows of Table~\ref{tab:coeff_nloew_zzH}) averaging over the whole available invariant-mass range.
The tree level results are the same for the two coefficients, owing to the scalar coupling of the Higgs to $\PZ$ bosons and to the $s$-channel topology of both lepton pairs. However, the numerical value depends rather strongly on $M_{\Pe^+\Pe^-}$ regions, especially when one of the two lepton-pairs is far from the $\PZ$ pole.
The inclusion of NLO EW corrections gives a small change in both coefficients around the on-shell region, while a more sizeable modification is found going towards the off-shell regime.
When comparing the two considered reference frames, it can be noticed that 
for $\alpha^{(1)}_{2,0}$ the NLO EW results deviate from each other for ($35\GeV<M_{\Pe^+\Pe^-}<85\GeV$, while for $\alpha^{(3)}_{2,0}$ the NLO EW results differ in the region $M_{\Pe^+\Pe^-}<45\GeV$. Looking at the overall trend of NLO EW results in Fig.~\ref{fig:MassDependence}, it is again clear that the 4$\ell$-CM helicity frame is better suited than the H-rest one, as the results are closer to the LO both in normalisation and in shape.

In conclusion, including NLO EW corrections changes the structure of the spin-density matrix and makes the standard projection method according to Eq.~\ref{eq:V1V2llll} not well defined in general. For some angular coefficients the LO values are not sizeably changed by EW radiative effects,
while for others the NLO corrections are dramatically large.

\subsection{Selection cuts and neutrino reconstruction}\label{sec:selectioncuts}
In the previous sections we have shown the manifold effects of the inclusion of higher-order corrections and off-shell effects on the extraction of angular coefficients from di-boson events, leading to potentially wrong conclusions when using the extracted coefficients to build observables sensitive to boson-pair spin correlation and entanglement.
The last, but definitely not the least, realistic effect that we would like to scrutinise is represented by selection cuts applied to boson-decay products. 

We come back to the angular structure of the EW-boson decay into two leptons introduced in Eq.~\ref{eq:V2ll}.
If $\pt{\ell},\eta_\ell$ cuts are applied on decay leptons, 
we write a set of coefficients $\{\tilde{\alpha}_{lm}\}$ in the following way,
\begin{equation}\label{eq:fid}
    \int_{-1}^{+1} \!\rd\cos\theta\int_{0}^{2\pi}\!\rd\phi\,\,\frac{\rd\sigma}{\rd\phi 
    \,\rd\cos\theta\,\rd \mc X}\, f^{(\mc X)}_{\rm cut}(\theta,\phi)\,\,Y_{lm}(\theta,\phi) = \frac{\rd \sigma}{\rd \mc X}\,\tilde{\alpha}_{lm}(\mc X)\,,
\end{equation}
where $f^{(\mc X)}_{\rm cut}(\theta,\phi)$ parametrises the application of cuts on the inclusive phase space. In formulas, 
the action of $f^{(\mc X)}_{\rm cut}(\theta,\phi)$ is defined by,
\begin{align}\label{eq:geoacceptance}
&\int_{\rm inc} \!\rd\pt{\ell}\,\rd\eta_\ell\, h(\pt{\ell},\eta_\ell)
\Theta(\pt{\ell}-\pt{\rm cut})
\Theta(\eta_{\ell}^{\rm (cut)}-|\eta_{\ell}|) \,=\\
=& \int_{-1}^{1} \!\rd\cos\theta\int_{0}^{2\pi}\!\rd\phi \,\,
 h(\pt{\ell}(\cos\theta,\phi),\eta_\ell(\cos\theta,\phi))\,
 \,
 \frac{\partial(\pt{\ell},\eta_\ell)}{\partial(\cos\theta,\phi)}\,\,f^{(\mc X)}_{\rm cut}(\cos\theta,\phi)\,,\nnb     
\end{align}
where $h(\pt{\ell},\eta_\ell)$ is a generic function of the lepton kinematics.
Since only the set of $\{\alpha_i(\mc X)\}$ coefficients allows for an interpretation in terms of spin polarisations and correlations, it is desirable to extract $\{\alpha_i(\mc X)\}$ directly from the data (or from simulated events). 
Expanding in terms of the first 8 spherical harmonics ($l\leq 2$) and weighting by the extracted $\{\tilde{\alpha}_i(\mc X)\}$ coefficients is known to be insufficient to reproduce correctly the correct shape of the cut unpolarised angular distributions \cite{Stirling:2012zt} and therefore, in general,
\beq
\tilde{\alpha}_{lm}(\mc X) \neq {\alpha}_{lm}(\mc X)\,.
\eeq
The same reasoning applies also to two-boson coefficents $\gamma_{l_1,m_1,l_2,m_2}$.

In Table~\ref{tab:coeff_nloqcd_inc_fid} we show how the standard fiducial cuts of Eq.~\ref{eq:ATLASfid} distort the 
values of most of angular coefficients, if the usual projection onto spherical harmonics with $l\leq 2$ is applied.
\begin{table}
  \begin{center}
    \begin{tabular}{crr}
    \hline
    \cellcolor{green!9} & \multicolumn{2}{c}{\cellcolor{green!9} ZZ} \\
      \hline\rule{0ex}{2.7ex}
       \cellcolor{blue!9}  
       & \cellcolor{blue!9} inclusive
       & \cellcolor{blue!9} ATLAS fiducial\\
       \hline\rule{0ex}{2.7ex}
$\alpha^{(1)}_{1,0}$ 	 &  $-0.00093 (5 )$ 	 & $  -0.00025 (7 )$ \\[0.1cm]
$\alpha^{(1)}_{2,0}$ 	 &  $0.02807 (5 )$ 	 & $  0.02075 (7 )$ \\[0.1cm]
$\alpha^{(3)}_{1,0}$ 	 &  $-0.00102 (5 )$ 	 & $  -0.00141 (7 )$ \\[0.1cm]
$\alpha^{(3)}_{2,0}$     &  $0.02794 (5 )$ 	 & $  0.02371 (7 )$ \\[0.1cm]
$\gamma_{1,0,1,0 }$ 	 &  $-0.00146 (2 )$ 	 & $  -0.00145 (2 )$ \\[0.1cm]
$\gamma_{2,0,2,0 }$ 	 &  $0.00167 (1 )$ 	 & $  0.00195 (2 )$ \\[0.1cm]
$\alpha^{(1)}_{2,-2 }$ 	 &  $-0.01000 (2 )$ 	 & $  -0.02614 (3 )$ \\[0.1cm]
$\alpha^{(2)}_{2,-2 }$ 	 &  $-0.01001 (2 )$ 	 & $  -0.02223 (4 )$ \\[0.1cm]
\hline
    \end{tabular}  
    \end{center}
  \caption{
  Angular coefficients at NLO QCD for off-shell $\PZ\PZ$ production at the LHC. All numbers have been computed integrating fully differential weights from {\scshape POWHEG-BOX-RES} \cite{Chiesa:2020ttl} multiplied by suitable combinations of spherical harmonics ($\ell\leq 2$) in the decay angles, according to Eq.~\ref{eq:coscos}. The inclusive setup defined in Eq.~\eqref{eq:inclsetup} (left) and the fiducial setup of Eq.~\eqref{eq:ATLASfid} (right) are understood. Numerical uncertainties are shown in parenthesis. A fixed renormalisation and factorisation scale $\mu_{\rm F}=\mu_{\rm R}=\MZ$ is assumed.
  \label{tab:coeff_nloqcd_inc_fid}}
\end{table}
In the shown results we include the dominant NLO QCD corrections, but similar results can be derived at LO and in the presence of EW corrections. 
For some coefficients, \emph{e.g.} $\gamma_{1,0,1,0}$, the effect of cuts is almost absent, while for others, both of single-boson and of two-boson kind, the effects of selection cuts is even beyond 100\%. 
This further confirms that the $l\leq 2$ expansion does not apply anymore in fiducial setups, even with rather loose selections. 
The $R_c$ correlation observable at NLO QCD in the presence of fiducial ATLAS cuts reads $R_c^{\rm fid}= 1.87(2)$, to be compared with the $R_c^{\rm inc}=1.73(2)$ obtained in the inclusive setup. The distortion is milder than expected, owing to partial cancellations between the numerator and denominator in Eq.~\ref{eq:Rcdef}.

The cut effects on coefficients associated to the Higgs-boson decay are weaker than in EW production. We consider a Higgs boson produced with finite transverse momentum at the LHC. We choose $\pt{\rm H}=25, \, 200\GeV$, in order to investigate both an inclusive and a boosted regime. In addition to the setup of Eq.~\ref{eq:mll10}, we apply ATLAS selections on charged leptons \cite{ATLAS:2019bsc},
\beq\label{eq:lepcuts}
\pt{\Pe}>7\GeV\,,\quad\pt{\mu}>5\GeV\,,\quad|y_{\Pe}|<2.47\,,\quad|y_{\mu}|<2.7\,.
\eeq
The numerical results at NLO EW accuracy are presented in Table~\ref{tab:cuts_nloew_zzH}.
\begin{table}
  \begin{center}
\begin{tabular}{lrrr}
\hline
\cellcolor{green!9}$\pt{\rm H}$ &\cellcolor{green!9} $25\GeV$ &\cellcolor{green!9} $25\GeV$ &\cellcolor{green!9} $200\GeV$ \\
\hline
\cellcolor{blue!9}lepton cuts &\cellcolor{blue!9} no cuts &
\cellcolor{blue!9}
Eq.~\ref{eq:lepcuts} &
\cellcolor{blue!9}
Eq.~\ref{eq:lepcuts} \\
\hline\rule{0ex}{2.7ex}\\[-0.4cm]
$\alpha^{(1)}_{2,0}$ 	 &   $-0.04511 (7 )$& $-0.07312 (8 )$ & $-0.05518 (8 )$\\[0.1cm]
$\alpha^{(3)}_{2,0}$	 &   $-0.04515 (7 )$& $-0.06498 (8 )$ & $-0.05336 (8 )$\\[0.1cm]
$\gamma_{1,0,1,0 }$ 	 &   $0.00011 (2 )$ & $0.00025 (2 )$ &  $0.00016 (2 )$\\[0.1cm]
$\gamma_{2,0,2,0 }$ 	 &   $0.01082 (2 )$ & $0.01229 (2 )$ &  $0.01134 (2 )$ \\[0.1cm]
$\gamma_{1,1,1-1 }$ 	 &   $-0.00047 (2 )$& $-0.00058 (3 )$ & $-0.00056 (3 )$  \\[0.1cm]
$\gamma_{1,-1,1,1 }$ 	 &   $-0.00047 (2 )$& $-0.00068 (3 )$ &  $-0.00055 (3 )$\\[0.1cm]
$\gamma_{2,1,2,-1 }$ 	 &   $-0.00779 (2 )$& $-0.00845 (3 )$ & $-0.00724 (2 )$  \\[0.1cm]
$\gamma_{2,-1,2,1 }$ 	 &   $-0.00783 (2 )$& $-0.00592 (2 )$ &  $-0.00788 (2 )$\\[0.1cm]
$\gamma_{2,2,2,-2 }$ 	 &   $0.00488 (2 )$ & $0.00549 (3 )$ &  $0.00523 (3 )$ \\[0.1cm]
$\gamma_{2,-2,2,2 }$ 	 &   $0.00489 (2 )$ & $0.00577 (3 )$  &  $0.00504 (3 )$\\[0.1cm]
\hline
\end{tabular}  
\end{center}
\caption{
 Angular coefficients at NLO EW for a SM Higgs-boson decay into four charged leptons (see Eq.~\eqref{eq:Higgsdec}) in the default setup of Eq.~\eqref{eq:mll10}, assuming a finite-$p_{\rm T}$ Higgs boson and including selections on the charged-leptons kinematics.
 All numbers have been computed integrating fully differential weights from {\scshape MG5\_aMC@NLO} \cite{Alwall:2014hca,Frederix:2018nkq} multiplied by suitable combinations of spherical harmonics ($\ell\leq 2$) in the decay angles, according to Eq.~\ref{eq:coscos}.
 Numerical uncertainties are shown in parentheses. The $4\ell$-CM helicity frame is understood.
  \label{tab:cuts_nloew_zzH}}
\end{table}
As found for inclusive $\PZ\PZ$ production, the selection cuts hamper the projection method,
leading to large, unphysical modifications to all considered coefficients. Owing to different selections for different lepton flavours, the symmetry relations of Eq.~\ref{eq:symmHiggs1} are also violated.
The moderate cut effects compared to those in Table~\ref{tab:coeff_nloqcd_inc_fid} can be traced back to the special structure of a narrow-width scalar resonance (the Higgs boson) which is boosted and produces two rather collinear vector bosons, while in EW production the two bosons are typically back to back. In addition, the weak bosons in the Higgs-boson rest frame decays are mostly longitudinally polarised and therefore the decay products are less sensitive to transverse-momentum cuts than in EW production, where the bosons are mostly transverse \cite{Denner:2021csi}.

The extraction of coefficients can be also hampered by neutrino reconstruction. This is the case for inclusive $\PW^\pm\PZ$ production at the LHC.
The standard techniques relying on on-shell requirements \cite{ATLAS:2019bsc}, as well as alternative machine-learning methods \cite{Grossi:2020orx}, do not enable to reconstruct properly the single-neutrino kinematics from $\PW$-boson decays, leading to bad distortions of the decay-angle distribution shapes, even in the absence of fiducial cuts.  
In Table~\ref{tab:vrec_inc_wz} we consider the inclusive setup of Eq.~\ref{eq:inclsetup} for $\PW^+\PZ$ production, comparing the coefficients computed with MC-truth neutrino kinematics with those obtained with reconstructed neutrino kinematics (according to the ATLAS on-shell technique \cite{ATLAS:2019bsc}). 
\begin{table}
  \begin{center}
    \begin{tabular}{crr}
    \hline
    \cellcolor{green!9} & \multicolumn{2}{c}{\cellcolor{green!9} $\PW^+$Z} \\
      \hline\rule{0ex}{2.7ex}
       \cellcolor{blue!9}  
       & \cellcolor{blue!9} MC-truth
       & \cellcolor{blue!9} $\nu$-reco\\
       \hline\rule{0ex}{2.7ex}
$\alpha^{(1)}_{1,0}$	 &  $-0.0464 (3 )$ 	 &  $-0.0222 (3 )$ \\[0.1cm]
$\alpha^{(1)}_{2,0}$		 &  $0.0283 (3 )$ 	 &  $0.0147 (3 )$ \\[0.1cm]
$\alpha^{(3)}_{1,0}$	 	 &  $0.0040 (3 )$ 	 &  $0.0010 (3 )$ \\[0.1cm]
$\alpha^{(3)}_{2,0}$	 	 &  $0.0298 (3 )$ 	 &  $0.0228 (3 )$ \\[0.1cm]
$\gamma_{1,0,1,0 }$ 	 &  $-0.0052 (1 )$ 	 &  $-0.0012 (1 )$ \\[0.1cm]
$\gamma_{2,0,2,0 }$ 	 &  $0.0013 (1 )$ 	 &  $-0.0013 (1 )$ \\[0.1cm]
$\alpha^{(1)}_{2,-2}$	 	 &  $-0.0120 (2 )$ 	 &  $-0.0002 (2 )$ \\[0.1cm]
$\alpha^{(3)}_{2,-2}$	 	 &  $-0.0074 (2 )$ 	 &  $-0.0023 (2 )$ \\[0.1cm]
\hline
    \end{tabular}  
    \end{center}
  \caption{
  Angular coefficients at NLO QCD for off-shell $\PW^+\PZ$ production at the LHC in the inclusive setup defined in Eq.~\eqref{eq:inclsetup}, with ($\nu$-reco) and without (MC-truth) neutrino reconstruction. All numbers have been computed integrating fully differential weights from {\scshape POWHEG-BOX-RES} \cite{Chiesa:2020ttl} multiplied by suitable combinations of spherical harmonics ($\ell\leq 2$) in the decay angles, according to Eq.~\ref{eq:coscos}. Numerical uncertainties are shown in parentheses. A fixed renormalisation and factorisation scale $\mu_{\rm F}=\mu_{\rm R}=\MZ$ is assumed.
  \label{tab:vrec_inc_wz}}
\end{table}
The reconstruction effects are remarkably large for almost all considered coefficients. Notice that the reconstruction does not only affect coefficients associated to the $\PW$ boson. In order to define decay angles starting from the di-boson CM frame (see discussion in Sect.~\ref{sec:AngCoeff}), the neutrino reconstruction is needed to access di-boson system, therefore affecting also the coefficients associated to the $\PZ$ boson.

\subsection{Extrapolation to the inclusive phase space}\label{sec:modeldep}
The results of Sect.~\ref{sec:selectioncuts} confirm that both selection cuts and neutrino reconstruction invalidate the projection method (see Eqs.~\ref{eq:fid} and \ref{eq:geoacceptance}). 
An extrapolation to the inclusive phase space is therefore unavoidable for a meaningful extraction of the angular coefficients.

For simplicity we focus on single-boson coefficients $\alpha_{l,m}$ in a generic LHC process. Thus, the needed extrapolation boils down to accessing the function $f^{(\mc X)}_{\rm cut}(\theta,\phi)$ introduced in Eq.~\ref{eq:fid}.
If one assumes that nature is SM-like, $f^{(\mc X)}_{\rm cut}(\theta,\phi)$ can be 
extracted via a numerical simulation of the angular distributions both in the inclusive phase space and after applying the wanted fiducial cuts. The ratio between such inclusive and fiducial distributions, up to an overall normalisation factor, represents the $f^{(\mc X)}_{\rm cut}(\theta,\phi)$ angular acceptance in the SM. It can be computed to the highest perturbative accuracy (in QCD and EW couplings), including parton-shower, hadronisation and detector effects, associating to it a suitable systematic uncertainty of both theoretical and experimental kind. Then it can be applied to signal events in the fiducial volume to obtain the angular distributions in the inclusive phase space.
This procedure is model dependent, as the effect of cuts and neutrino reconstruction can be different for SM distributions compared to those found in the presence of new-physics effects, leading to a model-dependent shape of $f^{(\mc X)}_{\rm cut}(\theta,\phi)$.

\begin{figure}
  \centering
    \includegraphics[scale=0.36]{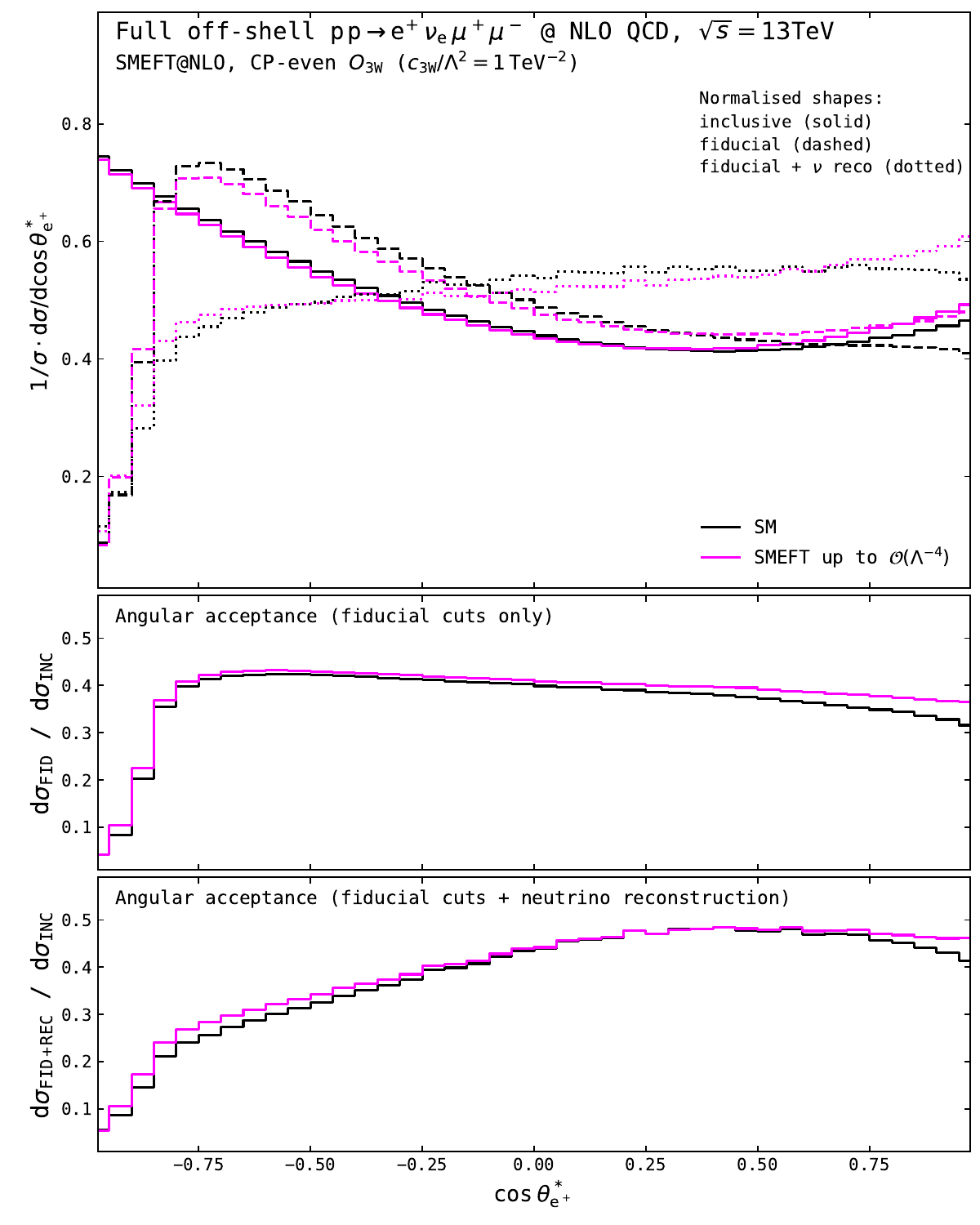}   
    \includegraphics[scale=0.36]{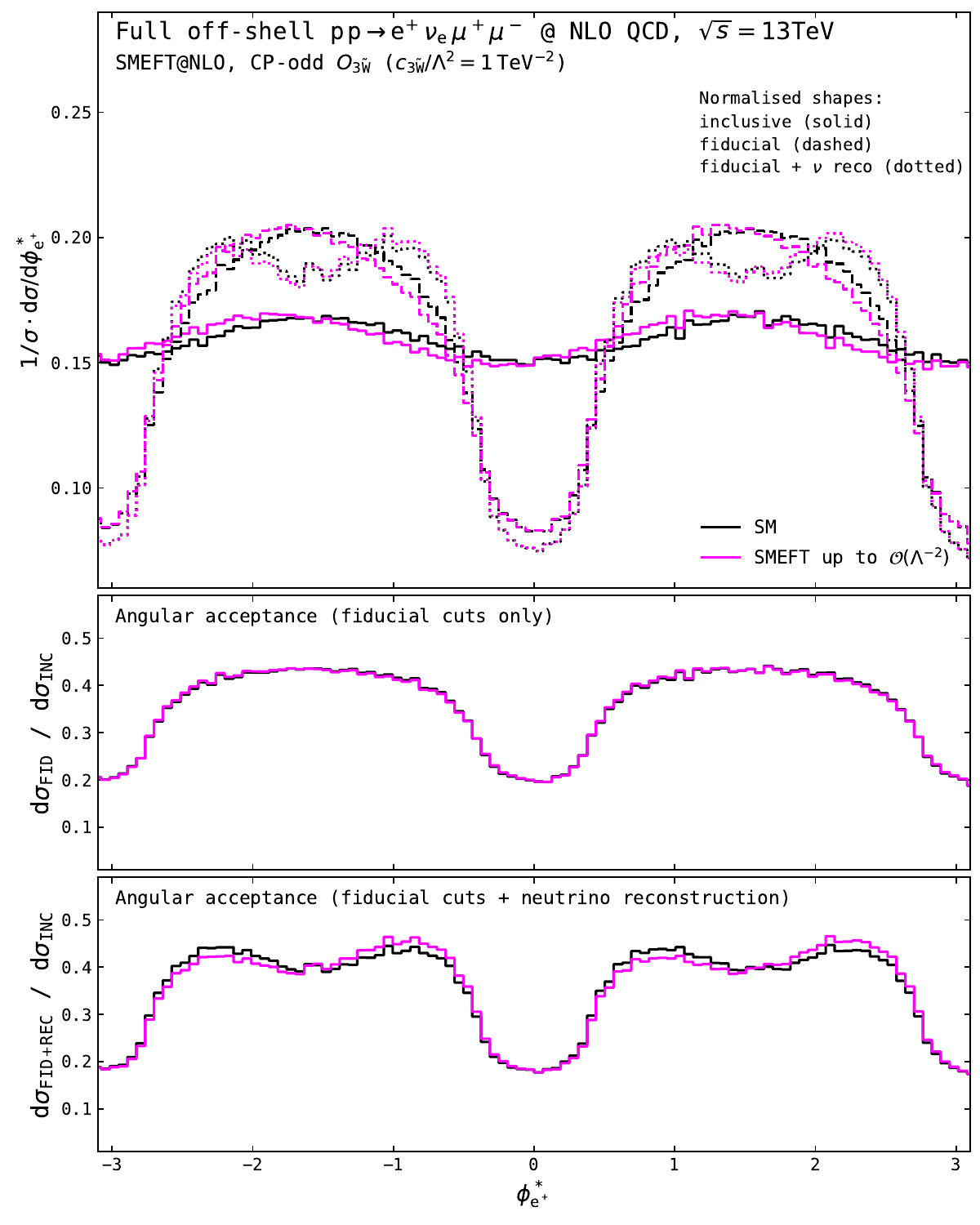}   
  \caption{Angular acceptance in the SM and in the SMEFT for a $\PW^+$-boson leptonic decay 
            in off-shell $\PW^+\PZ$ production at the LHC. Numerical results have been obtained at NLO QCD with MG5\_aMC@NLO \cite{Alwall:2014hca,Degrande:2020evl,ElFaham:2024uop}.
            Linear and quadratic effects from the dimension-six, CP-even operator $O_{3\PW}$ are considered for the polar angle (left).
           Linear effects from the dimension-six, CP-odd operator $O_{\Tilde{3\PW}}$ are considered for the azimuthal angle (right).
           Top panels: distributions (normalised to have unit integral) in the inclusive setup (solid), after applying ATLAS fiducial cuts \cite{ATLAS:2019bsc} (dashed), and neutrino reconstruction (dotted).
           Middle panels: ratio of fiducial distributions over inclusive ones (without neutrino reconstruction).
           Bottom panels: same as middle panels but with neutrino reconstruction.
  }\label{fig:SMEFTexample}
\end{figure}

To illustrate this aspect, we have computed angular distributions for $\PW^+\PZ$ production at NLO QCD accuracy, both in the SM and in the presence of dimension-six SMEFT effects modifying the triple-gauge coupling in the EW sector. In order to scrutinise both CP-even and CP-odd effects we have employed an extended version of the SMEFT@NLO UFO model \cite{Degrande:2020evl,ElFaham:2024uop}.
In Fig.~\ref{fig:SMEFTexample} we present the numerical results for polar- and azimuthal-angle distributions associated to the $\PW$-boson decay.
For the polar angle we have included both linear and quadratic effects coming from the 
CP-even operator $O_{3\PW}$, assuming a Wilson coefficient $c_{3\PW}/\Lambda^2=1{\rm TeV}^{-2}$.
The same value is chosen for the CP-odd operator $O_{3\Tilde{\PW}}$ which is considered at linear level in the case of the azimuthal-angle distribution.
Inclusively, the $O_{3\PW}$ operator is expected to enhance the right-handed component of the $\PW$ boson compared to the SM \cite{ElFaham:2024uop}, \emph{i.e.} increasing the number of events in the region $\cos\theta_1\approx 1$. The $O_{3\Tilde{\PW}}$ introduces an additional $\sin2\phi_1$ modulation in the azimuthal distribution, leading to a finite $\alpha^{(1)}_{2,2}$ coefficients (zero in the SM).
In the main panels of Fig.~\ref{fig:SMEFTexample} it can be seen how the inclusive distributions are heavily distorted by fiducial selections and neutrino reconstruction, leading to a depletion in the most populated region in the case of the polar angle, and to additional modulations in the azimuthal angle.
At a first glance, the angular acceptances do not look so different between SM and SMEFT, especially if the MC-truth neutrino kinematics is considered. Somewhat larger shape differences are visible when neutrino reconstruction is applied.
We have applied SM angular acceptances to the corresponding fiducial SMEFT distributions and then extracted the most relevant angular coefficients. The same QCD accuracy (NLO) is considered for both SM and SMEFT distributions.
In Table~\ref{tab:extr} we compare the results of this procedure with the known SMEFT results obtained in the inclusive phase space.
\begin{table}
  \begin{center}
    \begin{tabular}{lcccc}
    \hline
    \cellcolor{green!9} & \multicolumn{2}{c}{\cellcolor{green!9} CP-even $O_{3{\PW}}$} & \multicolumn{2}{c}{\cellcolor{green!9} CP-odd $O_{3\Tilde{\PW}}$}\\
      \hline\rule{0ex}{2.7ex}
  \cellcolor{blue!9}  &  \cellcolor{blue!9} $\alpha^{(1)}_{1,0}$ &  \cellcolor{blue!9} $\alpha^{(1)}_{2,0}$ &  \cellcolor{blue!9} $\alpha^{(1)}_{2,-2 }$ &  \cellcolor{blue!9} $\alpha^{(1)}_{2,2 }$\\[0.1cm]
    \hline\\[-0.3cm]
SM, truth    & $-0.0464  ( 1  )$& $0.0283  ( 1  )$& $ -0.0122  ( 2  )$ & $0.0001  ( 2  ) $\\[0.1cm]
SMEFT, truth    & $-0.0411  ( 1  )$ & $0.0313  ( 1  )$& $ -0.0120  ( 2  )$ & $-0.0066  ( 2  ) $ \\[0.1cm]
Extrapol., no $\nu$-reco.    & $-0.038  ( 2  )$& $0.039  ( 2  )$& $ -0.0116  ( 3  )$ & $ -0.0082  ( 3  )$ \\[0.1cm]
Extrapol., with $\nu$-reco.    & $-0.049  ( 2  )$& $0.041  ( 2  )$& $ -0.0119  ( 3  )$ & $0.0035  ( 3  ) $ \\[0.1cm]
\hline
    \end{tabular}  
    \end{center}
  \caption{
SM extrapolation applied to fiducial SMEFT angular distributions of Fig.~\ref{fig:SMEFTexample} with or without neutrino reconstruction. The coefficients are extracted with the usual projections according to Eq.~\ref{eq:V2ll}. \label{tab:extr}}
\end{table}
In the absence of neutrino reconstruction some coefficients are reproduced by less than $10\%$. This is the case of the $\alpha^{(1)}_{1,0}$ and $\alpha^{(1)}_{2,-2}$ coefficients.
In the most realistic situation where both fiducial cuts and neutrino reconstruction are accounted for, the expected SMEFT coefficients are badly reproduced by the extrapolation with up to 30\% discrepancies 
for polar coefficients, and even a wrong sign for $\alpha^{(1)}_{2,2}$.

In conclusion, assuming a SM-like angular acceptance and applying it to data which may embed new-physics effects could lead to wrong results for the angular coefficients. The SM extrapolation, relying on the highest accuracy available, leads to meaningful results only in experimental investigations aiming at confirming or excluding the SM. 
Pursuing a model-independent extrapolation to the inclusive phase space needs more refined techniques.
In particular, if one assumes SM dynamics in the boson decay while remaining agnostic w.r.t. potential new physics in the production mechanism, it is indispensable to compute angular acceptances separately for each helicity state of the intermediate boson. In fact, for a definite helicity state, an EW boson decays in the same manner whatever the production mechanism, therefore a polarised-template fit would allow to extract in a model-independent way the angular coefficients in the inclusive phase space.
The multiple calculations \cite{Ballestrero:2017bxn,BuarqueFranzosi:2019boy,Ballestrero:2019qoy,Ballestrero:2020qgv,Denner:2020bcz,Denner:2020eck,Poncelet:2021jmj,Denner:2021csi,Le:2022lrp,Le:2022ppa,Denner:2023ehn,Dao:2023pkl,Hoppe:2023uux,Pelliccioli:2023zpd,Javurkova:2024bwa,Denner:2024tlu,Dao:2024ffg}
of LHC processes with definite helicities for intermediate bosons already allow to pursue this objective, at least for what concerns polar coefficients, thanks to their direct relation to polarisation fractions (see Sect.~\ref{subsec:Rc}).
Carrying out the same procedure for azimuthal coefficients is more cumbersome, owing to differential distributions that are not positive definite.

\section{Conclusions and outlook}\label{sec:conclus}
Accessing the intricate spin structure in di-boson processes
is becoming a substantial part of the LHC physics program as it 
allows to probe the electroweak and scalar sectors of the Standard Model, but also 
enhances the sensitivity to new-physics effects.
Measuring polarisations and spin correlations is gaining further interest because of their deep
connection with quantum entanglement and Bell-inequality violation. 

It is known that the spin state associated to an electroweak-boson pair is 
encoded in the angular structures that characterise the decay products of the bosons.
An angular expansion in spherical harmonics up to rank-2 is known to describe
the system at tree level, in the case of two-body boson decays, and in the absence of kinematic
selections on individual decay products.
Extracting the coefficients of this expansion gives a direct access to the 
spin-density-matrix entries which then can be combined to construct
observables, dubbed quantum observables,
sensitive to entanglement and to the violation of Bell inequalities.

However, such a simple extraction is hampered by a number of effects
that introduce higher-rank spherical harmonics in the decay-angle expansion, 
and possibly invalidate the two-qutrits interpretation of the considered processes.
By means of a phenomenological analysis of di-boson systems at the LHC, we have scrutinised
such effects, focusing on the off-shell modeling of weak bosons, the inclusion of
higher-order corrections in the QCD and electroweak coupling, and realistic collider effects like fiducial-cut application and neutrino-kinematic reconstruction. 

The off-shell effects are found to be negligible in phase-space regions which
are dominated by two on-shell intermediate weak bosons, while they become 
crucial for the interpretation of angular coefficients associated to non-resonant
production mechanisms. This is especially true for boson pairs produced in the decay
of a Standard-Model Higgs boson.

The inclusion of NLO QCD corrections in the case of leptonic decays of weak bosons
does not introduce higher-rank contributions, but sizeably modifies the helicity structure
in the production mechanism, leading to a change in the numerical value of several coefficients.
Despite the power counting, the impact of NLO EW corrections is not much smaller, in particular for $l=1$ spin-correlation coefficients. The inclusion of NLO EW corrections is especially 
relevant in the case of the Higgs-boson decay. 

For the latter case we have provided an estimate at both LO and NLO EW accuracy 
for two quantities known to provide sensitivity to entanglement and to
the violation of the CGLMP Bell inequality, respectively. While for both quantum observables the tree-level description is sufficient to claim entanglement and Bell-inequality violation, the inclusion of NLO EW corrections is relevant to estimate the size of such quantum effects and their interpretation e.g. in BSM models.

By means of a comparison performed in the Higgs-boson decay, we have also 
revived the importance of choosing a suitable reference frame for the quantisation of the boson spin, in order to enhance the sensitivity to entanglement and Bell non-locality.

The application of selection cuts and, when needed, the reconstruction of neutrino kinematics are part of the experimental measurement of the final-state leptons at the LHC, and constitute an obstacle to the faithful determination of the coefficients according to the tree-level angular expansion.
Our results confirm the presence of higher-rank spherical harmonics contributions, which in turn make it problematic to interpret the $l\leq 2$ coefficients in terms of spin states.
An extrapolation to the fully inclusive phase space is therefore essential to recover the correct interpretation of such quantities. However, we have shown that such a procedure, on top of being subject to theoretical and experimental systematic uncertainties, is not model independent.

Our results broaden the understanding of unavoidable effects that may
distort the tree-level picture of the spin-density matrix and consequently of the 
quantum observables. The improved theoretical grounds pave the way for 
refined experimental analyses and more complete interpretations of the 
spin structure of boson pairs produced at the LHC.

\section*{Acknowledgements}
We would like to thank Emidio Gabrielli, Uli Haisch, Fabio Maltoni, Antonio Mandarino and Marius Wiesemann for fruitful discussions. 
GP is grateful to Jakob Linder and Giulia Zanderighi for useful exchange on the {\scshape PowHeg-Box-Res} package for di-boson processes in the DPA \cite{Pelliccioli:2023zpd},
to Hesham El Faham and Eleni Vryonidou for providing assistance with the SMEFT numerical results of Ref.~\cite{ElFaham:2024uop}.
The authors acknowledge the COMETA COST Action CA22130 for financial support through the assignment of two Short-Term Scientific Missions (ref.'s 041bb009, cdc9af95) and one Dissemination Conference Grant (ref. 27f322e8).
MG and AV acknowledge the hospitality of the Max-Planck Institute for Physics in Garching (MPP) where the project implementation was initiated.

\bibliographystyle{JHEP}


\end{document}

%% file: macros.tex
\def\citere#1{\mbox{Ref.~\cite{#1}}}
\def\citeres#1{\mbox{Refs.~\cite{#1}}}

\newcommand{\newc}{\newcommand}
\newc{\nnb}{\nonumber}
\newc{\beqn}{\begin{eqnarray}}
\newc{\eeqn}{\end{eqnarray}}
\newc{\beq}{\begin{equation}}
\newc{\eeq}{\end{equation}}
\newc{\bit}{\begin{itemize}}
\newc{\eit}{\end{itemize}}
\newc{\ben}{\begin{enumerate}}
\newc{\een}{\end{enumerate}}
\newc{\bce}{\begin{center}}
\newc{\ece}{\end{center}}
\newc{\bfi}{\begin{figure}}
\newc{\efi}{\end{figure}}

\newcommand{\rd}{\mathrm d}

\newcommand{\rT}{{\mathrm{T}}}

\newcommand{\rL}{{\mathrm{L}}}

\newcommand{\eg}{\emph{e.g.}\ }
\newcommand{\mc}[1]{\ensuremath{\mathcal{#1}}}

\newcommand{\GeV}{\ensuremath{\,\text{GeV}}\xspace}

\usepackage{xspace}

\newcommand{\Pp}{\ensuremath{\text{p}}}
\newcommand{\Pe}{\ensuremath{\text{e}}\xspace}

\newcommand{\Pg}{\ensuremath{\text{g}}}

\newcommand{\PW}{\ensuremath{\text{W}}\xspace}
\newcommand{\PZ}{\ensuremath{\text{Z}}\xspace}

\newcommand{\MW}{\ensuremath{M_\PW}\xspace}

\newcommand{\MZ}{\ensuremath{M_\PZ}\xspace}

\newcommand{\pt}[1]{\ensuremath{p_{\text{T},#1}}\xspace}

\newcommand{\PB}{\textsc{Powheg-Box}}

\newcolumntype{.}{D{.}{.}{-1}}
\newcolumntype{d}[1]{D{.}{.}{#1}}
\colorlet{tableoverheadcolor}{gray!37.5}
\colorlet{tableheadcolor}{gray!25}
\colorlet{tablerowcolor}{gray!12.5}

\def\draftdate{\relax}
\def\mda{\relax}
\def\mua{\relax}
\def\mla{\relax}
\def\draft{
\def\thtystars{******************************}
\def\sixtystars{\thtystars\thtystars}
\typeout{}
\typeout{\sixtystars**}
\typeout{* Draft mode!
         For final version remove \protect\draft\space in source file *}
\typeout{\sixtystars**}
\typeout{}
\def\draftdate{\today}
\def\mua{\marginpar[\boldmath\hfil$\uparrow$]%
                   {\boldmath$\uparrow$\hfil}\color{black}%
                    \typeout{marginpar: $\uparrow$}\ignorespaces}
\def\mda{\color{red}\marginpar[\boldmath\hfil$\downarrow$]%
                   {\boldmath$\downarrow$\hfil}%
                    \typeout{marginpar: $\downarrow$}\ignorespaces}
\def\mla{\marginpar[\boldmath\hfil$\rightarrow$]%
                   {\boldmath$\leftarrow $\hfil}%
                    \typeout{marginpar: $\leftrightarrow$}\ignorespaces}
\def\Mua{\marginpar[\boldmath\hfil$\Uparrow$]%
                   {\boldmath$\Uparrow$\hfil}\color{black}%
                    \typeout{marginpar: $\uparrow$}\ignorespaces}
\def\Mda{\color{red}\marginpar[\boldmath\hfil$\Downarrow$]%
                   {\boldmath$\Downarrow$\hfil}%
                    \typeout{marginpar: $\downarrow$}\ignorespaces}
\def\Mla{\marginpar[\boldmath\hfil\textcolor{red}{$\Rightarrow$}]%
                   {\boldmath\textcolor{red}{$\Leftarrow $}\hfil}%
                    \typeout{marginpar: $\leftrightarrow$}\ignorespaces}
\overfullrule 5pt
\oddsidemargin 15mm
\marginparwidth 29mm
}

